\documentclass[12pt]{article}
\usepackage{amsmath,amssymb}
\usepackage{bm}
\usepackage{graphicx,color}
\usepackage{wrapfig}

\begin{document}

\begin{flushright}
KEK-TH-1894
\end{flushright}

\vspace{2cm}

\begin{center}

{\bf \LARGE  Probing the SUSY with  $10$ TeV stop mass \\
\vspace{0.2cm}
 in rare decays and CP violation of Kaon} 
\\

\vspace*{1.5cm}
{\large 
Morimitsu~Tanimoto $^{a}$
 and \ \
Kei~Yamamoto $^{b}$
} \\
\vspace*{1cm}

{\it
$^b$Department of Physics, Niigata University,\\
Niigata 950-2181, Japan\\
$^b$KEK Theory Center, IPNS, KEK,  \\
Tsukuba, Ibaraki 305-0801, Japan \\
}

\end{center}

\vspace*{0.5cm}

\begin{abstract}
{\normalsize 
 We probe the SUSY  at  the $10 \ {\rm TeV}$ scale
 in the rare decays and the CP violation of  the kaon.
 We focus on  the processes of $K_L \to \pi^0 \nu{\bar \nu}$ and $K^+ \to \pi^+ \nu{\bar \nu}$ 
combined with   the CP violating parameters $\epsilon_K$ and $\epsilon_K^\prime/\epsilon_K$.
The Z-penguin mediated by the chargino loop cannot enhance  $K_L \to \pi^0 \nu{\bar \nu}$ and
 $K^+ \to \pi^+ \nu{\bar \nu}$ 
because the left-right mixing of the stop is constrained by the  $125$ GeV Higgs mass. 
On the other hand, the Z-penguin mediated by the gluino loop can enhance 
the  branching ratios of both $K_L \to \pi^0 \nu  {\bar \nu}$ and
 $K^+ \to \pi^+ \nu  {\bar \nu}$. The former increases up to more than $1.0 \times 10^{-10}$, which is
 much larger than the SM prediction even if  the constraint of $\epsilon_K$ is imposed.
 It is remarkable that  the Z-penguin mediated by the gluino loop can
 enhance  simultaneously  $\epsilon^\prime_K/\epsilon_K$ and the branching ratio of
$K_L \to \pi^0 \nu  {\bar \nu}$, which increases  up to  $1.0 \times 10^{-10}$.
  We also study the decay rates of 
$K_L\rightarrow  \mu^+\mu^-$, $B^0\rightarrow \mu^+\mu^-$ and  $B_s\rightarrow \mu^+\mu^-$,
 which correlate  with the $K_L \to \pi^0 \nu  {\bar \nu}$ decay through the Z penguin.
 It is important to examine  the  $B^0\rightarrow \mu^+\mu^-$ process
since we expect the enough sensitivity of this decay mode  to the SUSY  at LHCb.
}
\end{abstract}

\newpage
\section{Introduction}
\label{sec:Intro}

The rare decays and the CP violation of the kaon  have given  us  important constraints
 for  new physics (NP) since the standard model (SM) contributions
are suppressed due to the flavor structure of the  Cabibbo-Kobayashi-Maskawa (CKM) matrix 
\cite{Cabibbo:1963yz,Kobayashi:1973fv}.
Typical examples are  the rare decay processes 
 $K_L \to \pi^0 \nu  {\bar \nu}$ and $K^+ \to \pi^+ \nu {\bar \nu}$, 
 which are clean theoretically \cite{Buras:1998raa,Buras:2015qea}.
These processes  have been  considered to be one of the powerful probes of NP
\cite{Bertolini:1986bs}-\cite{Hou:2014nna}.
In order to improve the previous experimental measurements \cite{Ahn:2009gb,Artamonov:2009sz},
new  experiments are going on.  One is the J-PARC KOTO experiment, which  is to  measure the decay rate of
 $K_L \to \pi^0 \nu  {\bar \nu}$ approaching to the SM predicted precision 
\cite{Togawa:2014gya,Shiomi:2014sfa}. Another one is 
 the CERN NA62 experiment 
to observe  the $K^+ \to \pi^+ \nu {\bar \nu}$ decay \cite{VenelinKozhuharovfortheNA62:2014lca}.

Especially,  the  $K_L \to \pi^0 \nu  {\bar \nu}$ process  is the CP violating one   
and provides the direct measurement of the CP violating phase in the CKM matrix. 
On the other hand, the  indirect CP violating parameter $\epsilon_K$, 
which is  induced by the $K^0-{\bar K^0}$ mixing, has given us the precise information
of the CP violating phase of the CKM matrix.
Another CP violating parameter $\epsilon^\prime_K/\epsilon_K$ was measured in the 
$K\rightarrow \pi\pi$ decay.
Therefore,  the $K_L \to \pi^0 \nu  {\bar \nu}$ process is expected  
to open the NP window in the CP violation 
by combining with $\epsilon_K$ and $\epsilon^\prime_K/\epsilon_K$.

 The $K_L \to \pi^0 \nu  {\bar \nu}$ and $K^+ \to \pi^+ \nu {\bar \nu}$ 
decays are dominated by the Z-penguin process, which is the flavor
changing neutral current (FCNC) through  loop diagrams. 
The   Z-penguin process also gives the large contribution to  $\epsilon^\prime_K/\epsilon_K$
due to the enhancement of the  $\Delta I=1/2$ amplitude \cite{Buras:2015yba}.
Actually, it  cancels  the dominant QCD penguin contribution significantly in the SM 
since it has the opposite sign to the QCD penguin amplitude.
On the other hand,  $\epsilon_K$ is given by the box diagram.
We expect the deviation from the SM prediction  with correlating among 
$K_L \to \pi^0 \nu  {\bar \nu}$, $K^+ \to \pi^+ \nu {\bar \nu}$,   
$\epsilon_K$ and $\epsilon^\prime_K/\epsilon_K$
due to the NP effect.
Furthermore, there may be  other correlations of  NP with  
the kaon rare decay $K_L\rightarrow  \mu^+\mu^-$ \cite{Buras:2015jaq}
and the $B$ meson rare decays  $B^0\rightarrow \mu^+\mu^-$, $B_s\rightarrow \mu^+\mu^-$,
which have been observed in the LHCb and CMS experiments \cite{Aaij:2012nna,CMS:2014xfa}
since the  Z-penguin process also contributes to these processes. 

 In this work, we discuss the minimal supersymmetric SM (MSSM) as the typical  NP.
  The recent searches for SUSY particles  at the LHC give us important constraints.
Since the lower bounds of masses of the SUSY particles  increase gradually, 
 the gluino mass is supposed to be  beyond  the scale of $2$ TeV
 \cite{Aad:2014wea,Chatrchyan:2014lfa,Aad:2014kra,ATLAS:2016kts,CMS:2016mwj}.
 The SUSY models have been also seriously constrained  by the Higgs discovery, in which
the Higgs mass is  $125$ GeV~\cite{Aad:2012tfa,Chatrchyan:2012ufa}. 
 These facts suggest a class of SUSY models with heavy sfermions.
If  the squark masses
are expected to be  ${\cal O}(10)$ TeV,
 the lightest Higgs mass can be pushed up to $125$ GeV \cite{Draper:2011aa},
  whereas all SUSY particles  can be out of the reach of the LHC experiment.
Therefore, the indirect search of the SUSY particles becomes important in the low energy
flavor physics \cite{Altmannshofer:2013lfa,Moroi:2013sfa,Tanimoto:2014eva}.
We discuss the CP violation related phenomena such as
 $K_L \to \pi^0 \nu  {\bar \nu}$, $K^+ \to \pi^+ \nu {\bar \nu}$,   
$\epsilon_K$ and $\epsilon^\prime_K/\epsilon_K$ in the framework of the high-scale SUSY 
with ${\cal O}(10)$ TeV. 
 
We can also consider the SUSY model with   the split-family \cite{Endo:2010fk,Ibe:2013oha}
 in which the third family of squarks/sleptons is heavy,  ${\cal O}(10)$ TeV,
while the first and second  ones of squarks/sleptons and the gauginos 
have relatively low masses ${\cal O}(1)$ TeV.
This model is motivated by the Nambu-Goldstone  hypothesis for
quarks and leptons in the first two generations \cite{Mandal:2010hc}. 
Although  there is no signals of the SUSY particles in the LHC experiment at present, 
this scenario  is not conflict with the present bound of the SUSY particles.
The split-family model is consistent with  the $125$ GeV Higgs mass
 \cite{Aad:2012tfa,Chatrchyan:2012ufa} and the muon $g-2$ \cite{Bennett:2006fi}.
The stop mass with   ${\cal O}(10)$ TeV
 pushes  up the lightest Higgs mass  to $125$ GeV \cite{Draper:2011aa}. 
 The deviation from the SM prediction of the muon $g-2$ \cite{Hagiwara:2011af,Davier:2010nc}
 is explained by the   slepton 
 of the first and second family with the mass less than $1$ TeV \cite{Ibe:2013oha}. 
 Therefore, it is important to examine the split-family model  
in the rare decays and the CP violation at the low energy 
 \cite{Altmannshofer:2013lfa,Moroi:2013sfa,Tanimoto:2014eva}  as well as the direct search at LHC.

For many years, the rare decays and the CP violation in the $K$ and $B$ mesons 
have been successfully understood within the framework of the SM, 
where the source of the CP violation is the Kobayashi-Maskawa (KM) phase \cite{Kobayashi:1973fv}. 
On the other hand, there are new sources of the CP violation if the SM is 
extended to the SUSY model.
 For example, the soft squark mass matrices contain 
the CP violating phases, which contribute to  FCNC with the CP violation \cite{Gabbiani:1996hi}. 
Therefore, one expects to discover the SUSY contribution in the CP violating phenomena at the low energy. 
Actually,  we have found that the SUSY contribution  could be up to  $40\%$
 in the observed  $\epsilon_K$, but,
 it is minor in the CP violation of the $B$ meson at the high-scale of $10-50$ TeV 
\cite{Tanimoto:2014eva}. 
 Moreover, we have also found the sizable   contribution 
of  the high-scale SUSY 
to $K_L \to \pi^0 \nu  {\bar \nu}$ and $K^+ \to \pi^+ \nu {\bar \nu}$ 
in the non-minimal flavor violation (non-MFV) scenario  \cite{Tanimoto:2015ota}.

It is also important to take into account of $\epsilon^\prime_K/\epsilon_K$
because  the SM has potential difficulties
in describing the data for $\epsilon^\prime_K/\epsilon_K$ \cite{Buras:2015yba}. 
Therefore, we study  $\epsilon^\prime_K/\epsilon_K$ 
in the  SUSY model with the non-MFV scenario.
We  discuss $K_L \to \pi^0 \nu  {\bar \nu}$ and $K^+ \to \pi^+ \nu {\bar \nu}$
with  the CP violations, $\epsilon_K$ and $\epsilon^\prime_K/\epsilon_K$
 in the framework of the  SUSY at ${\cal O}(10)$ TeV.
 In addition, we discuss the SUSY contribution to   the decay processes 
$K_L\rightarrow  \mu^+\mu^-$,  $B^0\rightarrow \mu^+\mu^-$  and $B_s\rightarrow \mu^+\mu^-$.

We  have already presented the numerical predictions of
the branching ratios  $K_L \to \pi^0 \nu  {\bar \nu}$ and $K^+ \to \pi^+ \nu {\bar \nu}$ in ref.\cite{Tanimoto:2015ota},
where all squarks/sleptons and the gauginos are  at ${\cal O}(10)$ TeV.  
However,  those numerical results should be revised with the ones of this paper 
since the relevant constraints are not imposed enough there.
In this paper, we also reexamine them comprehensively
 by taking account of the  gluino contribution as well as the chargino one with the large left-right mixing angle of squarks.

Our paper is organized as follows.
In section 2, we discuss the formulation of the rare decays, 
 $K_L \to \pi^0 \nu  {\bar \nu}$, $K^+ \to \pi^+ \nu {\bar \nu}$, $K_L\rightarrow \mu^+\mu^-$,
 $B^0\rightarrow \mu^+\mu^-$ and  $B_s\rightarrow \mu^+\mu^-$,
 and CP violations of  $\epsilon_K$ and $\epsilon^\prime_K/\epsilon_K$.
Section 3  gives our set-up of  the SUSY with  the $10$ TeV squark masses. 
In Sec.4, we present   our numerical results.
Sec.5 is devoted to the  summary and discussions. 
The relevant formulae are presented 
 in Appendices A, B and  C.

\section{Observables}

\subsection{$K_L \to \pi^0 \nu  {\bar \nu}$ and $K^+ \to \pi^+ \nu {\bar \nu}$}

Let us begin to discuss the kaon rare decays,
$K_L \to \pi^0 \nu  {\bar \nu}$ and  $K^+ \to \pi^+ \nu {\bar \nu}$, which are dominated  
 by the Z-penguin process  in the SM.
 In the estimation of the branching ratios of $K \to \pi \nu{\bar \nu}$, the hadronic matrix elements can be extracted with the isospin symmetry relation \cite{Marciano:1996wy,Mescia:2007kn}.  
These processes are theoretically clean because the long-distance contributions  are small
\cite{Buras:2004uu}, and then the theoretical uncertainty is estimated below several percent.
The accurate measurements of these  decay processes  provide the crucial tests of the  SM.  Especially, the $K_L \to \pi^0 \nu  {\bar \nu}$ process is purely the CP violating one, which can reveal the source of  the CP violating phase.
The basic formulae are presented in Appendix C1.
The SM predictions have been  discussed  by some works
 \cite{Buras:2015qea,Brod:2010hi,Blanke:2016bhf}.  They are given as
 \footnote{In our calculation, we use the CKM elements in the study of 
 the so-called universal unitarity triangle including the data of  the CP asymmetry $S_{J / \psi K_S}$ and the mass differences of $B$ mesons  without inputting $\epsilon_K$ (Strategy S1 in ref.\cite{Blanke:2016bhf} ). 
In this case, the SM prediction for $K \to \pi \nu {\bar \nu}$ shifts lower.
}:
 \begin{align}
{\rm BR}(K_L \to \pi^0 \nu  {\bar \nu})_{\rm SM}
&=(3.36 \pm 0.05)\times 10^{-11} \cdot
\left[ \frac{|V_{ub}|}{3.88 \times 10^{-3}} \right]^2
\left[ \frac{|V_{cb}|}{40.7 \times 10^{-3}} \right]^2
\left[ \frac{{\rm sin}(\gamma)}{{\rm sin}(73.2 ^{\circ})} \right]^2
 ,  \\
{\rm BR}(K^+ \to \pi^+ \nu {\bar \nu})_{\rm SM}
&=(8.39 \pm 0.30)\times 10^{-11} \cdot
\left[ \frac{|V_{cb}|}{40.7 \times 10^{-3}} \right]^{2.8}
\left[ \frac{\gamma}{73.2 ^{\circ}} \right]^{0.74}.
\label{SMprediction}
\end{align}

On the experimental side, the upper bound of the branching ratio of
 $K_L \to \pi^0 \nu  {\bar \nu}$ is given  by the KEK E391a experiment \cite{Ahn:2009gb},
and  the branching ratio of $K^+ \to \pi^+ \nu {\bar \nu}$  was measured by the BNL E787 and E949 experiments as follows \cite{Artamonov:2009sz}: 
\begin{align}
{\rm BR}(K_L \to \pi^0 \nu  {\bar \nu})_{\rm exp}
< 2.6 \times 10^{-8} ~(90\% {\rm C.L.}), \ \
{\rm BR}(K^+ \to \pi^+ \nu {\bar \nu})_{\rm exp}
=(1.73^{+1.15}_{-1.05})\times 10^{-10}  . 
\end{align}
At present, the J-PARC KOTO experiment is an in-flight measurement of
 $K_L \to \pi^0 \nu  {\bar \nu}$ approaching to the SM predicted precision 
\cite{Togawa:2014gya,Shiomi:2014sfa}, 
while the CERN NA62 experiment \cite{VenelinKozhuharovfortheNA62:2014lca}
 is expected for the precise measurement of the $K^+ \to \pi^+ \nu {\bar \nu}$ decay.
 
 The SUSY contribution has been studied in many works 
\cite{Colangelo:1998pm,Nir:1997tf,Buras:1997ij,Buras:1999da,Buras:2004qb,Isidori:2006qy}.
The sizable enhancement of these kaon decays was expected 
 through the large  left-right mixing of the chargino interaction
 in   $s_L \tilde t_i \chi^-$ and $d_L \tilde t_i \chi^-$ 
 at the SUSY scale of ${\cal O}(1)$ TeV \cite{Colangelo:1998pm,Buras:2004qb}. 
  We find that even at the ${\cal O}(10)$ TeV scale,  these decays are enhanced
   through the Z-penguin mediated by the gluino with  the large left-right mixing. 

\subsection{$\epsilon_K$}

Let us discuss another CP violating parameter $\epsilon_K$, 
which was  measured precisely.
Its hadronic matrix element  $\hat{B}_K$ is reliably determined by
the  lattice calculations as follows  \cite{Bae:2013lja,Aoki:2013ldr} :  
\begin{equation}
\hat B_K= 0.766 \pm 0.010\ .
\label{BK}
\end{equation}
Another theoretical uncertainty of $\epsilon_K$ is also reduced 
by removing the QCD correction factor of the two charm box diagram \cite{Ligeti:2016qpi}.
Thus, the  accurate estimate of the SM contribution  enables us to search for  NP such as SUSY.
The non-negligible SUSY contribution has been  expected  in $\epsilon_K$ even
 at the scale of ${\mathcal O}(100)$ TeV \cite{Altmannshofer:2013lfa,Moroi:2013sfa,Tanimoto:2014eva}.
 Consequently,  $\epsilon_K$ gives us one of the most important constraints 
 to predict the SUSY contribution in the  $K \to \pi \nu {\bar \nu}$ decays.
 In our calculation of $\epsilon_K$, 
 we investigate the SUSY contributions for  the box diagram, which is correlated with
   the $K_L \to \pi^0 \nu {\bar \nu}$ process directly.


\subsection{$\epsilon^\prime_K/\epsilon_K$}
 
The direct  CP violation  $\epsilon^\prime_K/\epsilon_K$  is also important to constrain the NP.
The basic formula for $\epsilon^\prime_K/\epsilon_K$  is given as follows
 \cite{Buras:2015yba,Bosch:1999wr,Buras:2000qz}:
\begin{equation}
	\frac{\epsilon_K^\prime}{\epsilon_K}	
	={\rm Im} (V_{td}V_{ts}^* \cdot F_{\epsilon'} )
\end{equation}
where 
\begin{equation}
	F_{\epsilon'}
	=P_0 + P_X X + P_Y Y + P_Z Z + P_E E \ ,
\end{equation}
with
\begin{equation}
X = C - 4 B^{(u)}, \ \ \ Y = C - B^{(d)}, \ \ \ Z= C + \frac{1}{4}D.  
\end{equation}
Functions $B$, $C$, $D$ and $E$ denote the loop-functions including SM and SUSY effects, 
which come from boxes with external $d {\bar d}$$(B^{(d)})$,
 $u{\bar u}$$(B^{(u)})$, Z-penguin($C$), photon-penguin($D$) and 
 gluon-penguins($E$).  
The coefficients $P_i$ are given by
\begin{equation}
P_i = r_i^{(0)} + r_i^{(6)} R_6 +r_i^{(8)} R_8,  
\end{equation} 
with the non-perturbative parameters $B_6^{(1/2)}$ and $B_8^{(3/2)}$ defined as 
\begin{equation}
R_6 \equiv B_6^{(1/2)} \left[ \frac{114.54{\rm MeV}}{m_s(m_c)+m_d(m_c)} \right]^2, \ \ \
R_8 \equiv B_8^{(3/2)} \left[ \frac{114.54{\rm MeV}}{m_s(m_c)+m_d(m_c)} \right]^2.
\end{equation} 
The numerical values of $r_i^{(0,8,6)}$ are presented in \cite{Buras:2015yba}.

The most important parameters to predict  $\epsilon^\prime_K/\epsilon_K$  are
the non-perturbative parameters $B_6^{(1/2)}$ and $B_8^{(3/2)}$.
Recently, the RBC-UKQCD lattice collaboration \cite{Blum:2015ywa,Bai:2015nea} gives 
\begin{equation}
 B_6^{(1/2)}=0.57\pm 0.15, \qquad   B_8^{(3/2)} =0.76\pm 0.05,
 \label{B6B8}
\end{equation} 
which predict  $(\epsilon^\prime_K/\epsilon_K)_{\rm SM}=(1.9\pm 4.5)\times 10^{-4}$ 
in the SM \cite{Buras:2015yba}.
This SM prediction is much smaller than the experimental result \cite{PDG}
\begin{equation}
(\epsilon^\prime_K/\epsilon_K)_{\rm exp}=(16.6\pm 2.3)\times 10^{-4} . 
\end{equation}
This disagreement between the SM prediction and the experimental value may suggest
  NP in the kaon system, however there are several open questions that
have to be answered to conclude it \cite{Buras:2015yba}.
We use these values of $B_6^{(1/2)}$ and $B_8^{(3/2)}$ with $3\sigma$ in our calculation.

The dominant contribution to Z penguin, $C$,  comes from chargino mediated one
and gluino mediated one if the large left-right mixing of squarks is allowed.
On the other hand, the effect of  neutralino are suppressed 
\cite{Colangelo:1998pm}-\cite{Buras:1997ij}.
The chargino mediated Z-penguin $C({\chi^\pm})$
and the gluino mediated Z-penguin $C({\tilde g})$ 
are given as follows:
\begin{align}
V_{td}V_{ts}^*C({\chi^\pm})&= \frac{1}{8}\left( \frac{4 m_W^2}{g_2^2} \right) [P_{\rm ZL}^{sd}(\chi^{\pm})^*+
\frac{c_w^2}{s_w^2} P_{\rm ZR}^{sd}(\chi^{\pm})^*] , \\
V_{td}V_{ts}^* C({\tilde g})&= \frac{1}{8}\left( \frac{4 m_W^2}{g_2^2} \right)[P_{\rm ZL}^{sd}(\tilde g)^*+
\frac{c_w^2}{s_w^2}P_{\rm ZR}^{sd}(\tilde g)^*],
\end{align}
where $c_w^2=\cos ^2 \theta_W$ and $s_w^2=\sin^2 \theta_W$ with the Weinberg angle $\theta_W$, 
and the Z-penguin amplitudes  $P_{\rm ZL(R)}^{sd}(\chi^{\pm})$ and 
$P_{\rm ZL(R)}^{sd}(\tilde g)$ are  given in Eqs. (\ref{PLZ})  and  (\ref{PLG})   in Appendix B.


The box diagram effect is  suppressed compared with the  penguin diagram
if the SUSY-breaking scale $M_S$ satisfies $M_S \gg m_W$ \cite{Buras:1999da}.
Thus, the dominant SUSY contribution to $\epsilon^\prime_K/\epsilon_K$  is   given by 
the  Z-penguin mediated by the chargino and gluino.
Therefore, we should consider the correlation between  $\epsilon^\prime_K/\epsilon_K$
and the branching ratio of $K_L \to \pi^0 \nu  {\bar \nu}$.

Let us write  $\epsilon^\prime_K/\epsilon_K$ as
\begin{equation}
\left (\frac{\epsilon'_K}{\epsilon_K} \right )= \left (\frac{\epsilon'_K}{\epsilon_K} \right )_{\rm SM}+
\left (\frac{\epsilon'_K}{\epsilon_K} \right )_{Z}^L +
\left (\frac{\epsilon'_K}{\epsilon_K} \right )_{Z}^R \ ,
\end{equation}
where the second and the third terms denote the Z-penguin induced by the left-handed and
right-handed interactions of SUSY, respectively.
The both contributions are written  as follows \cite{Buras:2015jaq} :
\begin{equation}
\left (\frac{\epsilon'_K}{\epsilon_K} \right )^{L}_Z+\left (\frac{\epsilon'_K}{\epsilon_K} \right )^{R}_Z= 
-2.64\times 10^3 B_8^{(3/2)} 
 \left [ {\rm Im} \Delta_L^{sd}(Z)  + 
\frac{c_w^2}{s_w^2} {\rm Im} \Delta_R^{sd}(Z) \right ]
 \ ,
\end{equation}
where
\begin{equation}
\Delta_{L(R)}^{sd}(Z) = \frac{g_2}{8\pi^2 c_w}\frac{m_W^2}{2} P_{\rm ZL(R)}^{sd} \ .
\end{equation}

In order to see the correlation between
$\epsilon^\prime_K/\epsilon_K$  and  the $K_L \to \pi^0 \nu  {\bar \nu}$ decay,
it is helpful to write down the $K_L \to \pi^0 \nu  {\bar \nu}$ amplitude
 induced by  the chargino and gluino mediated Z-penguin 
in terms of  $\Delta_{L(R)}^{sd}(Z)$ as follows:
\begin{equation}
A(K_L \to \pi^0 \nu  {\bar \nu})_{{Z}} \sim  \left [ {\rm Im} \Delta_L^{sd}(Z)  + 
 {\rm Im} \Delta_R^{sd}(Z) \right ] \ ,
\end{equation}
as seen in  Appendix C1.

The  Z-penguin amplitude mediated by the chargino  dominates  the  left-handed coupling of the Z boson. Therefore, the chargino contribution  to  $\epsilon^\prime_K/\epsilon_K$
is opposite to  $K_L \to \pi^0 \nu  {\bar \nu}$. 
If the  Z-penguin mediated by the chargino enhances  $\epsilon^\prime_K/\epsilon_K$,
 the $K_L \to \pi^0 \nu  {\bar \nu}$ decay is suppressed considerably.
On the other hand, the Z-penguin amplitude  mediated by the gluino 
 gives the equal left-handed and right-handed Z couplings.
Then,  the  right-handed  Z coupling of the Z-penguin amplitude  is by a factor of $c_w^2/s_w^2\simeq 3.3$ larger than the left-handed one.
Therefore,  we can obtain the SUSY contribution 
which can enhance  simultaneously  $\epsilon^\prime_K/\epsilon_K$ and the branching ratio for
$K_L \to \pi^0 \nu  {\bar \nu}$. 
Actually, by choosing ${\rm Im}\Delta_{L}^{sd}(Z)>0$ and ${\rm Im}\Delta_{R}^{sd}(Z)<0$,
the region of  
\begin{equation}
|{\rm Im}\Delta_{R}^{sd}(Z)| < {\rm Im}\Delta_{L}^{sd}(Z) <3.3 |{\rm Im}\Delta_{R}^{sd}(Z) | \ ,
\label{buras-condition}
\end{equation}
can   enhance both $\epsilon^\prime_K/\epsilon_K$ and the branching ratio for
$K_L \to \pi^0 \nu  {\bar \nu}$. 
We discuss this case in our  numerical results.

\subsection{ $K_L\rightarrow \mu^+\mu^-$,
$B^0\rightarrow \mu^+\mu^-$ and $B_s\rightarrow \mu^+\mu^-$ decays}

The Z penguin also contributes  $K_L\rightarrow \mu^+\mu^-$,
$B^0\rightarrow \mu^+\mu^-$ and $B_s\rightarrow \mu^+\mu^-$ decays.
 These  decay amplitudes are governed by the axial semileptonic operator $O_{10}$,
 which is occurred by the Z-penguin top-loop and the W box diagram in the SM.  
 Those general formulae are presented  in Appendix C2.
The CMS and LHCb Collaboration have observed the branching ratio for $B_s\rightarrow \mu^+ \mu^-$  
, and $B^0\rightarrow \mu^+ \mu^-$  is also measured \cite{CMS:2014xfa}:
\begin{equation}
{\rm BR}(B_s\rightarrow \mu^+ \mu^-)_{\rm exp}= (2.8^{+0.7}_{-0.6}) \times 10^{-9}, \qquad
{\rm BR}(B^0\rightarrow \mu^+ \mu^-)_{\rm exp}= (3.9^{+1.6}_{-1.4}) \times 10^{-10}.
\end{equation}
The SM predictions have been given as \cite{Bobeth:2013uxa},
\begin{equation}
{\rm BR}(B_s\rightarrow \mu^+ \mu^-)_{\rm SM}= (3.65\pm 0.23) \times 10^{-9}, \ \ \
{\rm BR}(B^0\rightarrow \mu^+ \mu^-)_{\rm SM}= (1.06\pm 0.09) \times 10^{-10}.
\end{equation}

 On the other hand, the long-distance effect is expected to be large
 in the $K_L\rightarrow \mu^+\mu^-$ process \cite{Isidori:2003ts}. Therefore,
 it may be difficult to extract the  effect of the Z-penguin process.
 The SM prediction of the short-distance contribution was given as  \cite{Buras:2015jaq},
\begin{equation}
{\rm BR}(K_L\rightarrow \mu^+ \mu^-)_{\rm SM}= (0.8\pm 0.1) \times 10^{-9}.
\end{equation}
The experimental data  of $K_L\rightarrow \mu^+ \mu^-$ is \cite{PDG}
\begin{equation}
{\rm BR}(K_L\rightarrow \mu^+ \mu^-)_{\rm exp}= (6.84\pm 0.11) \times 10^{-9},
\label{Kmumu}
\end{equation}
from which the constraint on the short-distance contribution has been estimated as \cite{Isidori:2003ts} :
\begin{equation}
{\rm BR}(K_L\rightarrow \mu^+ \mu^-)_{\rm SD} \le  2.5 \times 10^{-9}.
\label{KtomumuSDbound}
\end{equation}

Thus, the SUSY contribution  through the Z penguin is expected to be correlated 
among the rare decays of 
 $K_L \to \pi^0 \nu  {\bar \nu}$, $K^+ \to \pi^+ \nu {\bar \nu}$,
 $K_L\rightarrow \mu^+\mu^-$, $B^0\rightarrow \mu^+\mu^-$ and $B_s\rightarrow \mu^+\mu^-$
 as well as the CP violations of $\epsilon_K$ and  $\epsilon_K^\prime / \epsilon_K$.
 
\section{SUSY flavor mixing}

Recent LHC results for the SUSY search may suggest the high-scale SUSY, ${\mathcal O}(10-1000)$ TeV
\cite{Altmannshofer:2013lfa,Moroi:2013sfa,Tanimoto:2014eva}
since the lower bounds of the gluino mass and squark masses are close to  $2$ TeV.
Taking account of these recent results, we consider the possibility of the high-scale SUSY 
at $10 \ {\rm TeV}$, in which 
the $K \to \pi \nu {\bar \nu}$ decays and  $\epsilon^\prime_K/\epsilon_K$ with
 the constraint of  $\epsilon_K$ are discussed.
 
We also consider the split-family model, which has the specific spectrum of the SUSY particles
\cite{Endo:2010fk,Ibe:2013oha}.
This model is motivated by the Nambu-Goldstone  hypothesis for
quarks and leptons in the first two generations \cite{Mandal:2010hc}. 
 Therefore, the third family of squark/slepton is
 heavy,  for example, ${\cal O}(10)$ TeV while the first and second  family squarks/slepton have relatively low masses ${\cal O}(1)$ TeV. 
 The masses of bino and wino are assumed to be small 
close to the experimental lower bound, less than $1$ TeV.
The model was at first discussed in the $B_s-\bar B_s$ mixing \cite{Endo:2010fk}.
It explained successfully both  the $125$ GeV Higgs mass and the muon $g-2$ 
simultaneously \cite{Ibe:2013oha}.
The stop mass with   ${\cal O}(10)$ TeV pushes up the  Higgs mass  to $125$ GeV. 
 The deviation of the muon $g-2$ is explained by the   slepton 
 of the first and second familes with the mass less than $1$ TeV.
  Since the squark  masses of  the first and second families are also relatively low as well as the
sleptons, we expect the SUSY contribution in the kaon system becomes large.


The new flavor mixing and CP violation effect are induced through the quark-squark-gaugino and 
the lepton-slepton-gaugino couplings.
The $6\times 6$ squark mass matrix  $M_{\tilde q}^2$ in the super-CKM basis is diagonalized to  
the mass eigenstate basis in terms of the rotation matrix $\Gamma ^{(q)}$ as  
\begin{equation}
m_{\tilde q}^2=\Gamma^{(q)}M_{\tilde q}^2 \Gamma^{(q)  \dagger} \ ,
\end{equation}
where $\Gamma^{(q)}$ is $6\times 6$ unitary matrix, and  it is decomposed
into $3\times 6$ matrices as $\Gamma ^{(q)}=(\Gamma _{L}^{(q)}, \ \Gamma _{R}^{(q))})$.
The explicit matrix is shown in Appendix A.
We introduce twelve mixing parameters  $s_{12}^{qL,qR}$, $s_{23}^{qL,qR}$ and $s_{13}^{qL,qR}$, where $q=u, d$ for the squark mixing. In addition, we also introduce left-right (LR) mixing angles $\theta^{t, b}_{LR}$ .

In practice, we take  $s_{12}^{qL,qR}=0$, which is motivated by the almost degenerate squark masses of the first and the second  families
to protect the large contribution to the $K^0-\bar K^0$ mass difference $\Delta M_K$. 
 It is also known that the single mixing effect of  $s_{12}^{qL,qR}$
to $K \to \pi \nu {\bar \nu}$ is minor \cite{Colangelo:1998pm}.
Actually, we have  checked numerically that the contribution
 of  $s_{12}^{qL,qR}=0\sim 0.3$  is negligibly small.
There also appear the phases $\phi^{qL}_{ij}$ and  $\phi^{qR}_{ij}$
associated with the mixing angles, which bring new sources of the CP violations.
 In our work, we treat those mixing parameters and phases as free parameters
in the framework of the non-MFV scenario.

Since the Z-penguin processes give dominant contribution for 
 $K\rightarrow \pi\nu \bar\nu$ and $\epsilon^\prime_K/\epsilon_K$,
 we calculate the Z-penguin mediated by the chargino and gluino.
 The interaction is presented in  Appendix B.
 The relevant parameters are presented in the following section.

\section{Numerical analysis}
\label{sec:numerical}
\subsection{Set-up of parameters}

Let us discuss the decay rates of $K_L \to \pi^0 \nu  {\bar \nu}$ and
 $K^+ \to \pi^+ \nu {\bar \nu}$ processes
by choosing  a sample of the  mass spectrum in  
the high-scale SUSY model at  ${\cal O}(10)$ TeV.
The enhancements of these kaon rare decays  require
    the large left-right mixing with the large squark flavor mixing.
In order to show our results clearly,
we take a simple set-up for the high-scale SUSY model as follows:
\begin{itemize}
 \item We fix  the gluino, wino and bino  masses  $M_i (i=3,2,1)$ 
with   $\mu$ and
tan$\beta$ as:
   \begin{align}
   M_3=10\  {\rm TeV}, \quad M_2=3.3 \  {\rm TeV}, \quad M_1=1.6\  {\rm TeV}, \quad
\mu=10 \ {\rm TeV} , \quad
{\rm tan}\beta = 3,
   \label{stop}
   \end{align} 
   for the  high-scale SUSY. 
 \item We take the masses of  stop $\tilde t_1$, $\tilde t_2$,
and sbottom $\tilde b_1$, $\tilde b_2$  as a sample set
   \begin{align}
   m_{\tilde t_1}=10 \ {\rm TeV}, \quad m_{\tilde t_2}=15\  {\rm TeV} , \quad
   m_{\tilde b_1}=10 \ {\rm TeV}, \quad m_{\tilde b_2}=15\  {\rm TeV}.
   \label{stop}
   \end{align}
On the other hand,
 we take the masses of the first and second family up-type and down-type squarks 
 around $15$TeV within $5-15\%$ relevantly. 
 This mass spectrum of  the first and second families does not so change our numerical results
 because the third family squarks dominate the Z-penguin induced by the chargino and gluino 
interactions in our model.

\item 
 We take the   left-right mixing angles
 \begin{equation}
\theta^t_{LR}=0.07 \qquad  {\rm  and}  \qquad \theta^b_{LR}=0.1-0.3 \ ,
\end{equation}
where  $\theta^t_{LR}$  is estimated by input of  the stop masses in   Eq.(\ref{stop}) with
 the large $A$ term,  which is constrained by the $125$ GeV Higgs mass due to the large radiative correction
  \cite{Draper:2011aa}.
On the other hand, there is no  strong constraint for the left-right mixing of the down squarks 
from the $B$ meson experiments in the region of  ${\cal O}(10)$ TeV
\footnote{The metastability of vacuum  can also  constrain the left-right mixing
for the down squark sector \cite{Park:2010wf}.
In order to justify our set-up of the left-right mixing angle,   the more precise analysis of the vacuum stability 
is important. }.
Therefore, we take rather large values to see the enhancement of the $K_L \to \pi^0 \nu  {\bar \nu}$ decay.
\item 
The flavor mixing parameters $s_{ij}^{qL}$ and $s_{ij}^{qR}$ of
the up and down sectors are free parameters, 
and are  varied in
   \begin{align}
   s_{i3}^{uL} , s_{i3}^{dL}=0  \sim 0.3 \ (i=1,2) , 
   \qquad  s_{i3}^{uR}, s_{d3}^{dR} =0  \sim 0.3 \  (i=1,2), 
   \end{align}
   where the upper bound $0.3$ is given by the experimental constraint of 
   the $K^0-\bar K^0$ mass difference  $\Delta M_K$.
 As discussed in the previous section, we ignore the mixing between the first and second family of squarks, $s_{12}^{qL}$,
  and then, can avoid the large contribution from  $s_{12}^{qL}$ to $\Delta M_K$.
This single mixing effect of $s_{12}^{qL}$  to the Z-penguin mediated by the chargino is  known to be minor
 compared with double mixing effect
\cite{Colangelo:1998pm,Buras:2004qb}.
  Namely, the SUSY contributions of 
 the  $K_L \to \pi^0 \nu  {\bar \nu}$ and
 $K^+ \to \pi^+ \nu {\bar \nu}$ processes are  dominated 
by the   double mixing of the stop and sbottom.

\item 
The phase parameters $\phi^{qL(R)}_{13}$ and $\phi^{qL(R)}_{23}$ are also free parameters.
We scan them in $-\pi \sim \pi$ randomly.
 \item 
We  neglect the minor contribution from the slepton and sneutrino.
We also neglect the charged Higgs contribution, which is  tiny due to the CKM mixing.
 \item For  non-perturbative parameters $B_6^{(1/2)}$ and $B_8^{(3/2)}$,
which are key ones to estimate $\epsilon^\prime_K/\epsilon_K$, 
we use the RBC-UKQCD result
$B_6^{(1/2)}=0.57\pm 0.15$ and $B_8^{(3/2)} =0.76\pm 0.05$ in Eq.(\ref{B6B8}).
We scan them within the $3\sigma$ error-bar.
\end{itemize}
 We use  the CKM elements $|V_{cb}|$, $|V_{ub}|$, 
$|V_{td}|$ in ref.\cite{Blanke:2016bhf} with $3\sigma$ error bars,
   which are obtained in the framework of the SM. If there is a large SUSY contribution  
 to the kaon and the $B$ meson systems, the values of the CKM elements may be changed.  
Actually, the SUSY contribution 
is comparable to the SM one for  $\epsilon_K$  in our following numerical analyses,
however, very small for the CP violations and  the  mass differences of the $B$ mesons at the  ${\cal O}(10)$TeV scale
of squarks \cite{Tanimoto:2014eva}.
We use the CKM element in the study of 
 the unitarity triangle including the data of  the CP asymmetries and the mass differences of $B$
 mesons  without inputting $\epsilon_K$ (Strategy S1 in ref.\cite{Blanke:2016bhf} ). 
\subsection{Results in the  SUSY at $10$ TeV}
Let us discuss the case of the  high-scale SUSY,  where all squark/slepton are
at the $10$ TeV scale.

At first, we discuss the contribution of the Z-penguin induced by the chargino  
to the $K_L \to \pi^0 \nu  {\bar \nu}$  and $K^+ \to \pi^+ \nu  {\bar \nu}$ processes. 
In this case, the  left-right mixing of the up squark sector controls the magnitude
 of the Z-penguin amplitude.
Since the  $A$ term is considerably constrained  by the $125$ GeV Higgs mass,
the left-right mixing angle cannot be large in our mass spectrum, at most $\theta^t_{LR}=0.07$
 as presented in the above set-up.
Therefore, we cannot obtain the enhancement of  those processes
\footnote{If we take the smaller mass for $m_{\tilde t_2}$ in Eq.(\ref{stop}), for example, $12$ TeV, the left-right mixing angle
 can be chosen to be  larger than $0.1$.  However, the contribution of $m_{\tilde t_1}$  is canceled
 by the one of $m_{\tilde t_2}$ due to the small mass difference significantly.}.
Actually, the predicted branching ratios of 
$K_L \to \pi^0 \nu  {\bar \nu}$  and $K^+ \to \pi^+ \nu  {\bar \nu}$
 deviate from the prediction of the SM with  of order   $10\%$.
 Thus, we conclude that the Z-penguin mediated by chargino cannot 
 bring large enhancement for  
the $K_L \to \pi^0 \nu  {\bar \nu}$  and $K^+ \to \pi^+ \nu  {\bar \nu}$ decays
 due to the constraint of the $125$ GeV Higgs mass. 
This result is consistent with the recent work \cite{Endo:2016aws},
where the metastability of vacuum   constrains the left-right mixing for the up squark sector.

On the other hand, the Z-penguin induced by the gluino could be large
 due to the  large down-type left-right mixing $\theta^b_{LR}=0.1-0.3$.
In our set-up of parameters, 
 we show  the predicted branching ratios,  ${\rm BR}(K_L \to \pi^0 \nu  {\bar \nu})$ versus 
${\rm BR}(K^+ \to \pi^+ \nu  {\bar \nu})$ in fig.1,
where the mixing  $s_{13}^{dL,dR}$ and $s_{23}^{dL,dR}$  are scanned in $0-0.3$ 
and the left-right mixing angle $\theta_{LR}^b$ is fixed to be $0.3$.
Here the Grossman-Nir bound is shown by the slant green  line \cite{Grossman:1997sk}.
In order to see the $\theta^b_{LR}$ dependence, we also present the 
 ${\rm BR}(K_L \to \pi^0 \nu  {\bar \nu})$ versus 
${\rm BR}(K^+ \to \pi^+ \nu  {\bar \nu})$ in fig.2  and fig.3,
in which $\theta_{LR}^b=0.2$ and $0.1$ are fixed respectively.
As seen in figs.1-3, the branching ratio of ${\rm BR}(K_L \to \pi^0 \nu  {\bar \nu})$
depends on considerably the left-right mixing angle $\theta_{LR}^b$.
 The enhancement of  ${\rm BR}(K_L \to \pi^0 \nu  {\bar \nu})$
requires the left-right mixing angle to be larger than $0.1$.

\begin{figure}[h]
\begin{minipage}[]{0.48\linewidth}
\includegraphics[width=8cm]{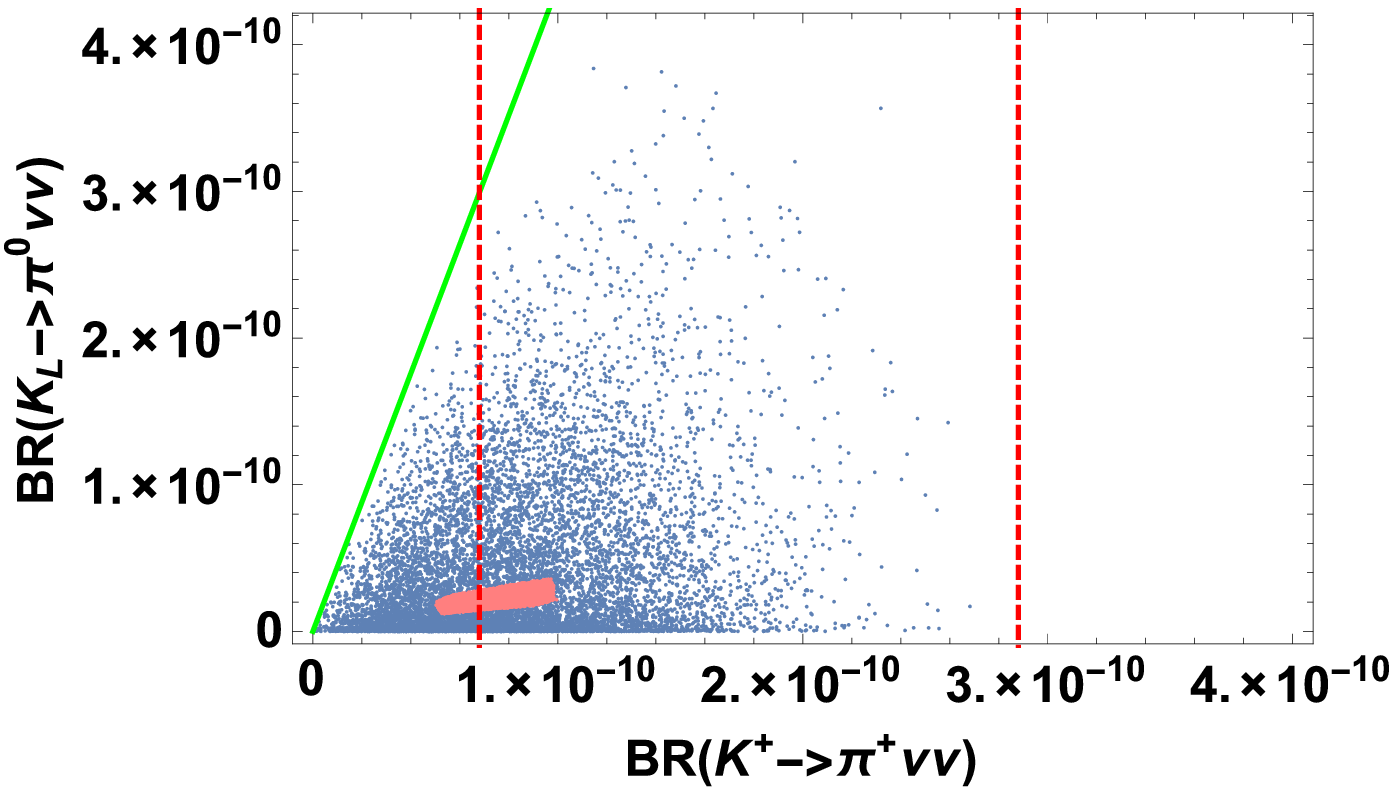}
\caption{The predicted region for ${\rm BR}(K_L \to \pi^0 \nu  {\bar \nu})$ versus 
${\rm BR}(K^+ \to \pi^+ \nu  {\bar \nu})$ without imposing $\epsilon_K$
where $\theta_{LR}^b=0.3$. 
The green line corresponds to  the Grossman-Nir bound.
The dashed red lines denote the $1\sigma$ experimental bounds for
 ${\rm BR}(K^+ \to \pi^+ \nu {\bar \nu})$.
The pink denotes the SM prediction.}
\end{minipage}
\hspace{5mm}
\begin{minipage}[]{0.48\linewidth}
\includegraphics[width=8cm]{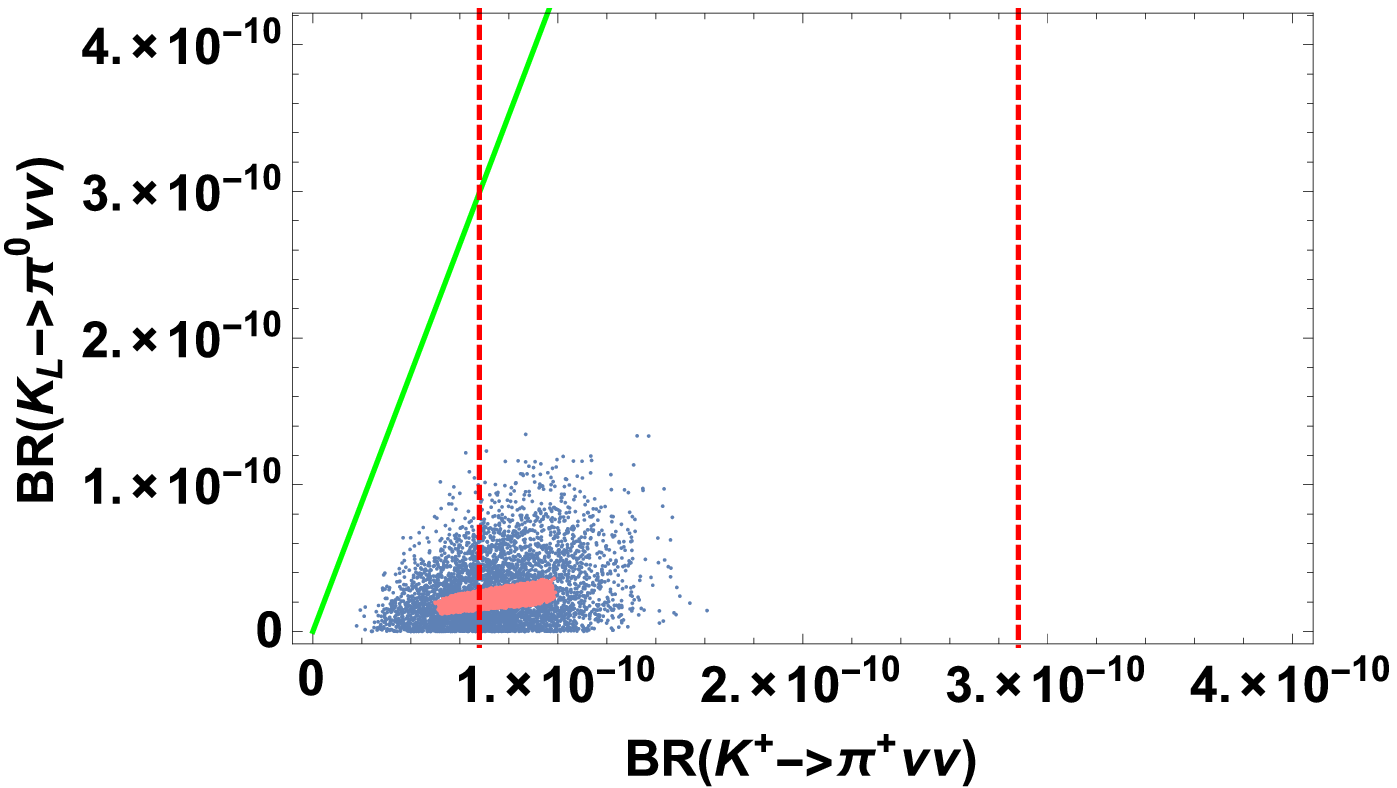}
\caption{
The predicted region for ${\rm BR}(K_L \to \pi^0 \nu  {\bar \nu})$ versus 
${\rm BR}(K^+ \to \pi^+ \nu  {\bar \nu})$,
without imposing $\epsilon_K$, where $\theta_{LR}^b=0.2$.
Notations are same as in Figure 1.
}
\end{minipage}
\end{figure}
\begin{figure}[h!]
\begin{minipage}[]{0.48\linewidth}
\includegraphics[width=8cm]{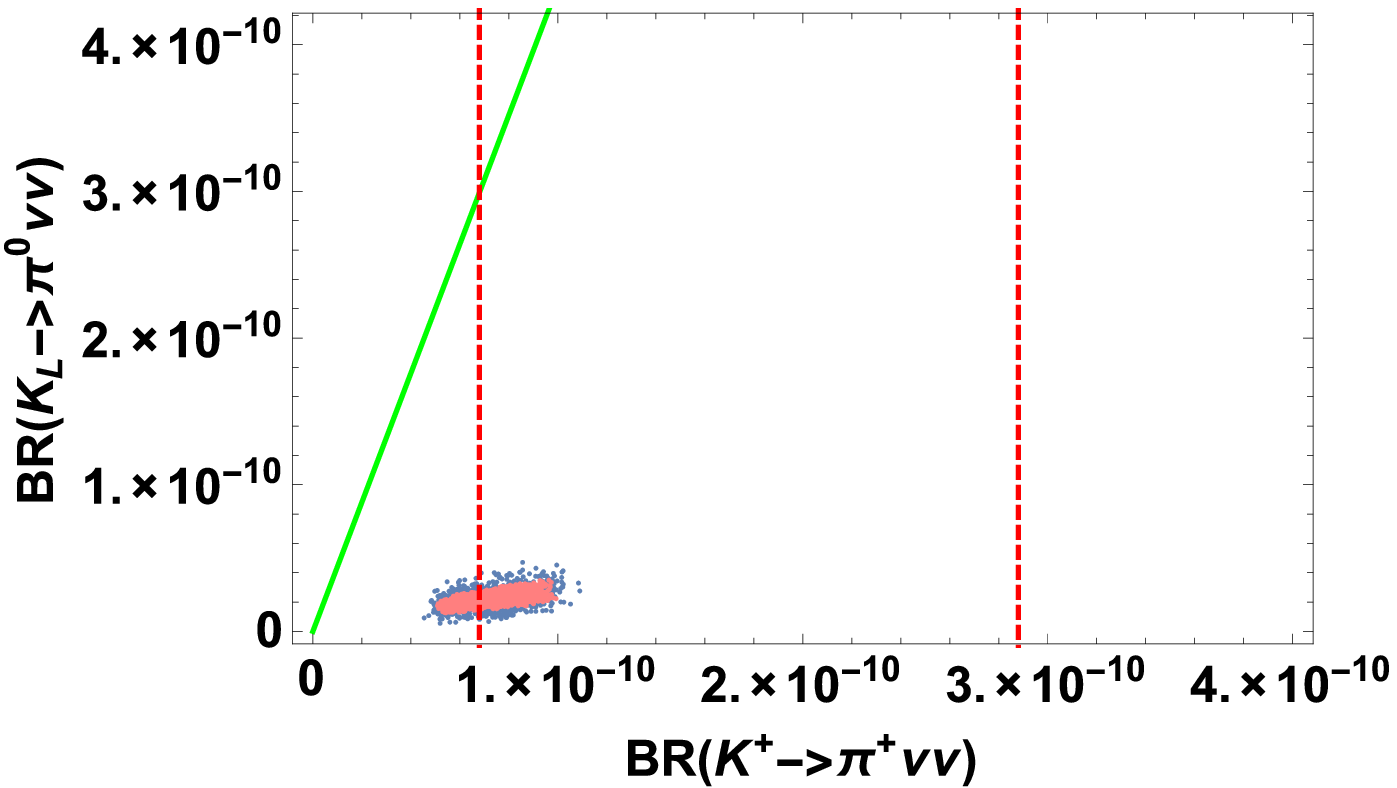}
\caption{
The predicted region for ${\rm BR}(K_L \to \pi^0 \nu  {\bar \nu})$ versus 
${\rm BR}(K^+ \to \pi^+ \nu  {\bar \nu})$,
without imposing $\epsilon_K$, where $\theta_{LR}^b=0.1$.
Notations are same as in Figure 1.
}
\end{minipage}
\hspace{5mm}
\begin{minipage}[]{0.48\linewidth}
\includegraphics[width=8cm]{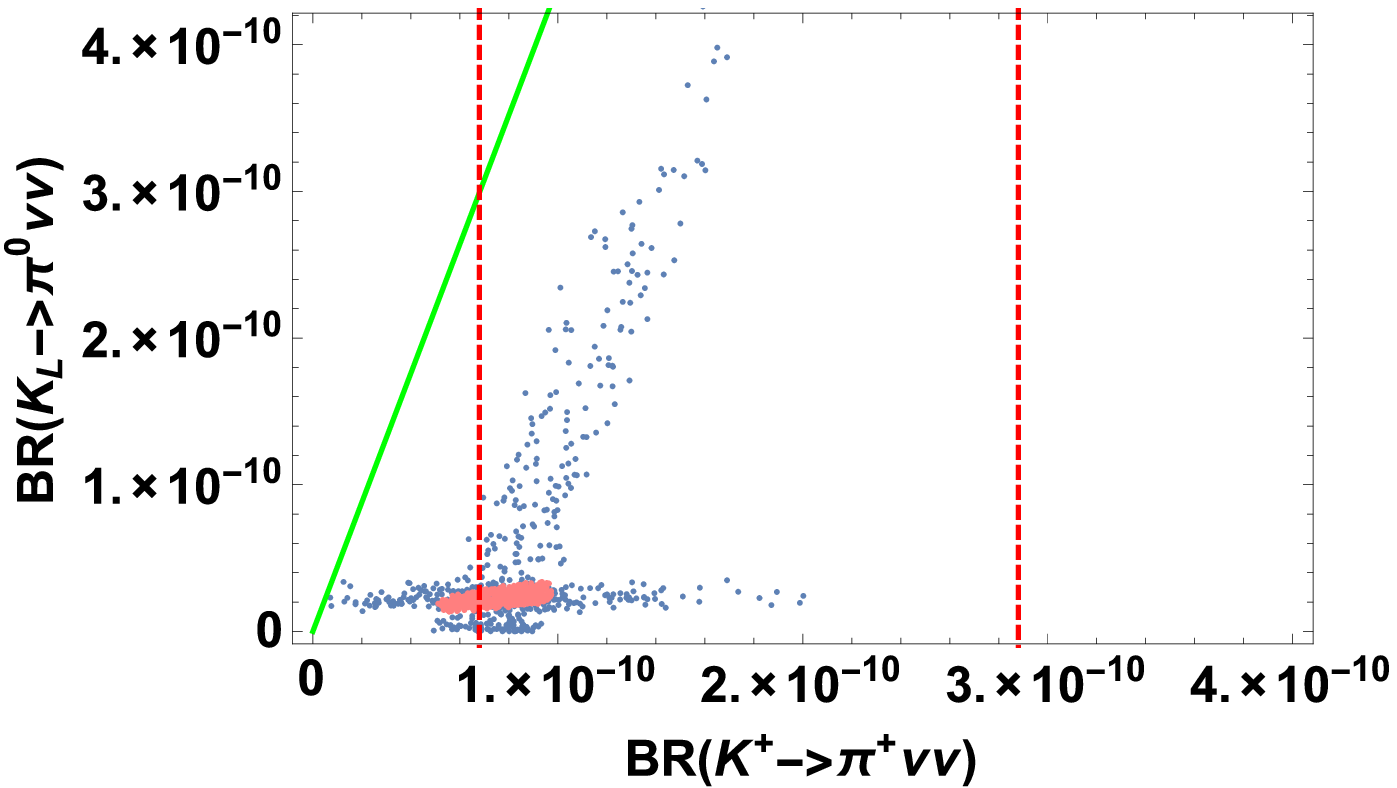}
\caption{
The predicted region for ${\rm BR}(K_L \to \pi^0 \nu  {\bar \nu})$ versus 
${\rm BR}(K^+ \to \pi^+ \nu  {\bar \nu})$,
with imposing $\epsilon_K$, where $\theta_{LR}^b=0.3$.
Notations are same as in Figure 1.
}
\end{minipage}
\end{figure}

Though the constraint of the experimental value of  $\epsilon_K$ is important, it is not imposed in figs.1-3.
Let us  take account of $\epsilon_K$.
The gluino contribution  to $\epsilon_K$ depends on the phase differences of  
$\phi^{dL(dR)}_{13}$ and $\phi^{dL(dR)}_{23}$, which are associated with flavor mixing angles.
In order to avoid the large contribution of the relatively light squarks to  $\epsilon_K$, 
the phases $\phi^{dL,dR}_{13}-\phi^{dL,dR}_{23}$ should be tuned near 
$n \times \pi/2 (n=-2,-1, 0,1,2)$ 
\footnote{The interpretation of the relation between the phase dependence of $K \to \pi \nu  {\bar \nu}$ and the one of $\epsilon_K$ was discussed in ref. \cite{Blanke:2009pq}.}.
For the  phase cycle in the  branching ratio,
${\rm BR}(K_L \to \pi^0 \nu  {\bar \nu})$ is a half of the one in  $\epsilon_K$.
Therefore, the  enhancement of  ${\rm BR}(K_L \to \pi^0 \nu {\bar \nu})$ is realized 
at $\phi^{dL}_{13}-\phi^{d}_{23} \simeq \pi/2$ and $\phi^{dR}_{13}-\phi^{dR}_{23} \simeq -\pi/2$, where
$\epsilon_K$ is enough suppressed. 
At $\phi^{dL}_{13}-\phi^{dL}_{23} \simeq -\pi/2$ and $\phi^{dR}_{13}-\phi^{dR}_{23} \simeq \pi/2$,
the  SUSY contribution to the $K_L \to \pi^0 \nu {\bar \nu}$ process is  the  opposite to the SM one, and then the branching ratio is suppressed compared with the SM prediction.

We show  the  predicted region for ${\rm BR}(K_L \to \pi^0 \nu  {\bar \nu})$ versus 
${\rm BR}(K^+ \to \pi^+ \nu  {\bar \nu})$,
with imposing $\epsilon_K$  where $\theta_{LR}^b=0.3$ is fixed in fig.4.
There are two direction in the predicted  plane of 
${\rm BR}(K_L \to \pi^0 \nu  {\bar \nu})$ versus 
${\rm BR}(K^+ \to \pi^+ \nu  {\bar \nu})$.
The direction of the enhancement of  ${\rm BR}(K_L \to \pi^0 \nu  {\bar \nu})$ corresponds to 
$\phi^{dL}_{13}-\phi^{dL}_{23} \simeq -\pi/2$ and $\phi^{dR}_{13}-\phi^{dR}_{23} \simeq \pi/2$,
and the enhancement of ${\rm BR}(K^+ \to \pi^+ \nu  {\bar \nu})$
 to  $\phi^{dL,dR}_{13}-\phi^{dL,dR}_{23} \simeq 0, \ \pi$.

As a result, it is found that ${\rm BR}(K_L \to \pi^0 \nu  {\bar \nu})$ can be enhanced up to 
$4 \times 10^{-10}$, which is much larger than the SM one, with satisfying the $\epsilon_K$ constraint.

We comment on
the constraint from $K^0-\bar K^0$ mass difference $\Delta M_K$.
 Our SUSY contribution of $\Delta M_K({\rm SUSY})$ is comparable with 
  the SM contribution $\Delta M_K({\rm SM})$. It is possible to fit the
  following condition keeping the enhancement of  ${\rm BR}(K_L \to \pi^0 \nu  {\bar \nu})$:
 \begin{equation}
 \frac{\Delta M_K}{\Delta M_K(SM)}= 0.75\sim 1.25 \ ,
 \end{equation}
 which is  the criterion of the allowed NP contribution in ref. \cite{Buras:2013ooa}.
 We also estimate the SUSY contribution to $\Delta M_{B^0}$ and $\Delta M_{B_s}$, 
which are  at most $10\%$ of the SM.

\begin{figure}[h]
\begin{minipage}[]{0.48\linewidth}
\includegraphics[width=8cm]{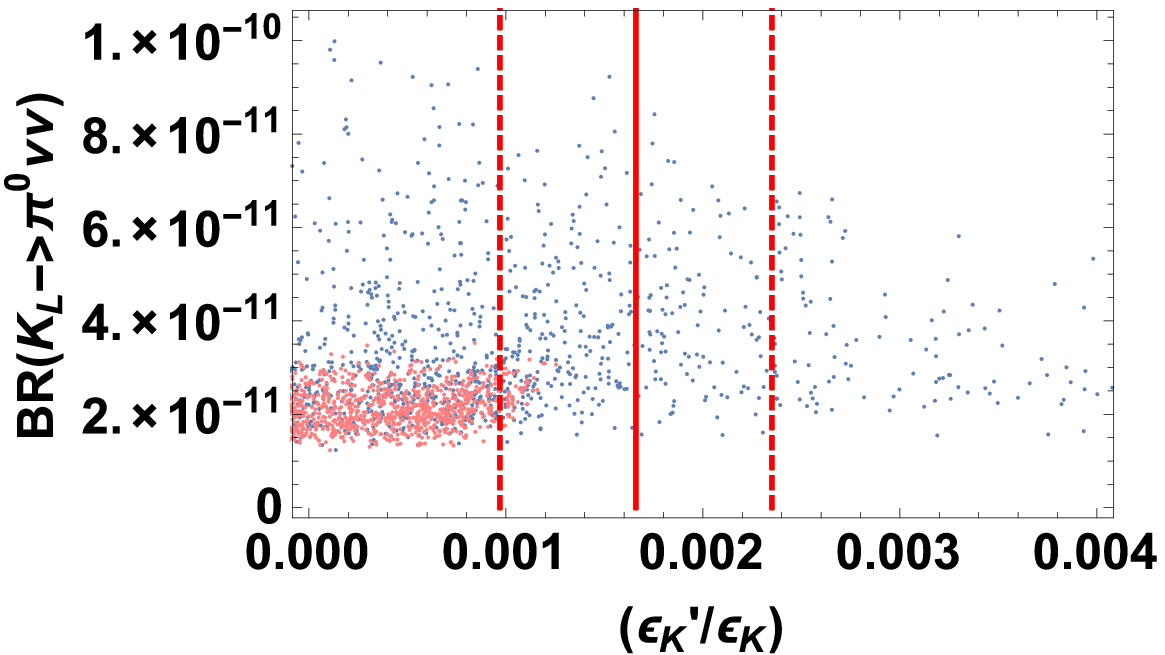}
\caption{The predicted 
${\rm BR}(K_L \to \pi^0 \nu  {\bar \nu})$ versus $\epsilon^\prime_K/\epsilon_K$,
 where the $Zsd$ coupling satisfies
the condition of  eq.(\ref{buras-condition}).
The vertical solid red line denotes the central value of the experimental data, 
and the dashed ones denote the  experimental bounds with $3\sigma$ for $\epsilon^\prime_K/\epsilon_K$.
The pink denotes the SM prediction.}
\end{minipage}
\hspace{5mm}
\begin{minipage}[]{0.48\linewidth}
\includegraphics[width=8cm]{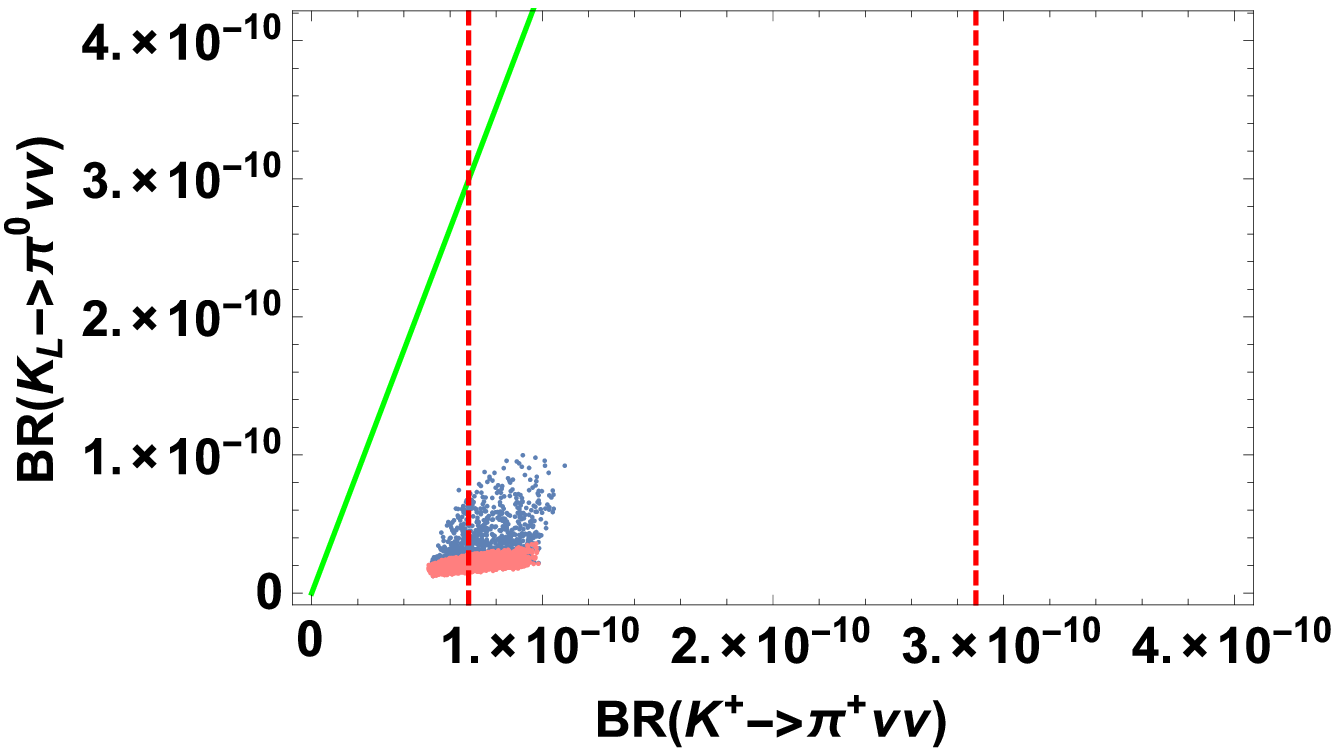}
\caption{The predicted region for ${\rm BR}(K_L \to \pi^0 \nu  {\bar \nu})$ versus 
${\rm BR}(K^+ \to \pi^+ \nu  {\bar \nu})$, 
where the $Zsd$ coupling satisfies the condition of  eq.(\ref{buras-condition}).
Notations are  same as in Figure 1.}
\end{minipage}
\end{figure}
Let us discuss the correlation between
${\rm BR}(K_L \to \pi^0 \nu  {\bar \nu})$ and $\epsilon^\prime_K/\epsilon_K$.
As discussed in subsection 2.3, both processes come from the imaginary part of the same Z-penguin,
and can be enhanced simultaneously once the condition Eq.(\ref{buras-condition}) is imposed.
In fig.5, we show the correlation between  ${\rm BR}(K_L \to \pi^0 \nu  {\bar \nu})$ and
$\epsilon^\prime_K/\epsilon_K$, where $Zsd$ coupling satisfies the condition 
of eq.(\ref{buras-condition}).
The constraint from $\epsilon_K$ is also imposed.
It is remarkable that  the Z-penguin mediated by the gluino  
 enhances  simultaneously  $\epsilon^\prime_K/\epsilon_K$ and the branching ratio for
$K_L \to \pi^0 \nu  {\bar \nu}$.  While the estimated $\epsilon^\prime_K/\epsilon_K$ fits the observed value,
 the branching ratio  of $K_L \to \pi^0 \nu  {\bar \nu}$ increases  up to  $1.0 \times 10^{-10}$.
In this region, the phase of Im$\Delta_L^{sd}$ and Im$\Delta_R^{sd}$ becomes opposite, 
so the enhanced region of ${\rm BR}(K_L \to \pi^0 \nu  {\bar \nu})$ is 
somewhat reduced by the cancellation between the left-handed coupling of Z
and  the right-handed one partially, compared with the result in  fig.4. 

The real part of $\Delta_L^{sd}$ and $\Delta_R^{sd}$ are small sufficiently 
 since $\phi^{dL,dR}_{13}-\phi^{dL,dR}_{23} \simeq \pm \pi/2$ are taken.
  Therefore, the SUSY contribution does not spoil the agreement of the real part of 
 the $K\to \pi\pi$ amplitude in the SM  with the experimental data.


In fig.6, we show the correlation between ${\rm BR}(K_L \to \pi^0 \nu  {\bar \nu})$ and
${\rm BR}(K^+ \to \pi^+ \nu  {\bar \nu})$.
In the parameter region where ${\rm BR}(K_L \to \pi^0 \nu  {\bar \nu})$ 
and $\epsilon^\prime_K/\epsilon_K$ are enhanced,  the branching ratio of $K^+ \to \pi^+ \nu  {\bar \nu}$ 
is not deviated from the SM.
It is understandable 
because $\phi^{dL,dR}_{13}-\phi^{dL,dR}_{23} \simeq \pm \pi/2$  is taken in order to enhance
 ${\rm BR}(K_L \to \pi^0 \nu  {\bar \nu})$ with the  $\epsilon_K$ constraint. 
On the other hand, ${\rm BR}(K^+ \to \pi^+ \nu  {\bar \nu})$ is dominated by 
the considerably sizable real part of the SM.  
The addition of the imaginary part of the SUSY contribution
does not change the SM prediction significantly.


 The Z-penguin process also contributes to another  
 kaon rare decay  $K_L\rightarrow  \mu^+\mu^-$,
and the $B$ meson rare decays,  $B^0\rightarrow \mu^+\mu^-$ and  $B_s\rightarrow \mu^+\mu^-$.
Therefore, we expect  them to correlate with the $K \to \pi \nu  {\bar \nu}$ decays.
In the $K_L\rightarrow  \mu^+\mu^-$ process, 
the long-distance effect is estimated to be large in ref. \cite{Isidori:2003ts}. Therefore,
  we only discuss the short-distance effect, which is dominated by the Z-penguin.
We show    ${\rm BR}(K_L \to \pi^0 \nu  {\bar \nu})$ versus
 ${\rm BR}(K_L\rightarrow  \mu^+\mu^-)$ in fig.7, where the constraint from $\epsilon_K$ is imposed.  
 It is noticed that the predicted value almost satisfies the bound for the short-distance contribution
  in Eq.(\ref{KtomumuSDbound}), presented as the red line.
 
 The clear correlation between two branching ratios is understandable because
  ${\rm BR}(K_L\rightarrow  \mu^+\mu^-)$ is sensitive only to the real part of Z-couplings.
When the enhancement of ${\rm BR}(K_L \to \pi^0 \nu  {\bar \nu})$ is found
  in the future, ${\rm BR}(K_L\rightarrow  \mu^+\mu^-)$ remains to be less than $10^{-9}$.
  On the other hand,  ${\rm BR}(K_L\rightarrow  \mu^+\mu^-)$ is larger than $10^{-9}$,
  there is no enhancement of ${\rm BR}(K_L \to \pi^0 \nu  {\bar \nu})$.
  This relation is testable in the future experiments.

 We  also show ${\rm BR}(K_L \to \pi^0 \nu  {\bar \nu})$ versus
   ${\rm BR} (B^0\rightarrow \mu^+\mu^-)$ in fig.8.
   We can expect the enhancement of  ${\rm BR} (B^0\rightarrow \mu^+\mu^-)$ in our set-up even if 
      ${\rm BR}(K_L \to \pi^0 \nu  {\bar \nu})$ is comparable to the SM one.
  Since LHCb will observe the  ${\rm BR} (B^0\rightarrow \mu^+\mu^-)$ \cite{Butler:2013kdw},
  this result is the attractive one in our model.

   On the other hand, 
  we do not see the correlation between  ${\rm BR}(K_L \to \pi^0 \nu  {\bar \nu})$ and
   ${\rm BR} (B_s\rightarrow \mu^+\mu^-)$ since the SM component of  ${\rm BR} (B_s\rightarrow \mu^+\mu^-)$
   is relatively large compared with  $B^0\rightarrow \mu^+\mu^-$.
 The enhancement of the $K_L \to \pi^0 \nu  {\bar \nu}$ decay rate
  is still consistent with the present experimental
 data of ${\rm BR} (B_s\rightarrow \mu^+\mu^-)$.
 
\begin{figure}[t!]
\begin{minipage}[]{0.48\linewidth}
\includegraphics[width=8cm]{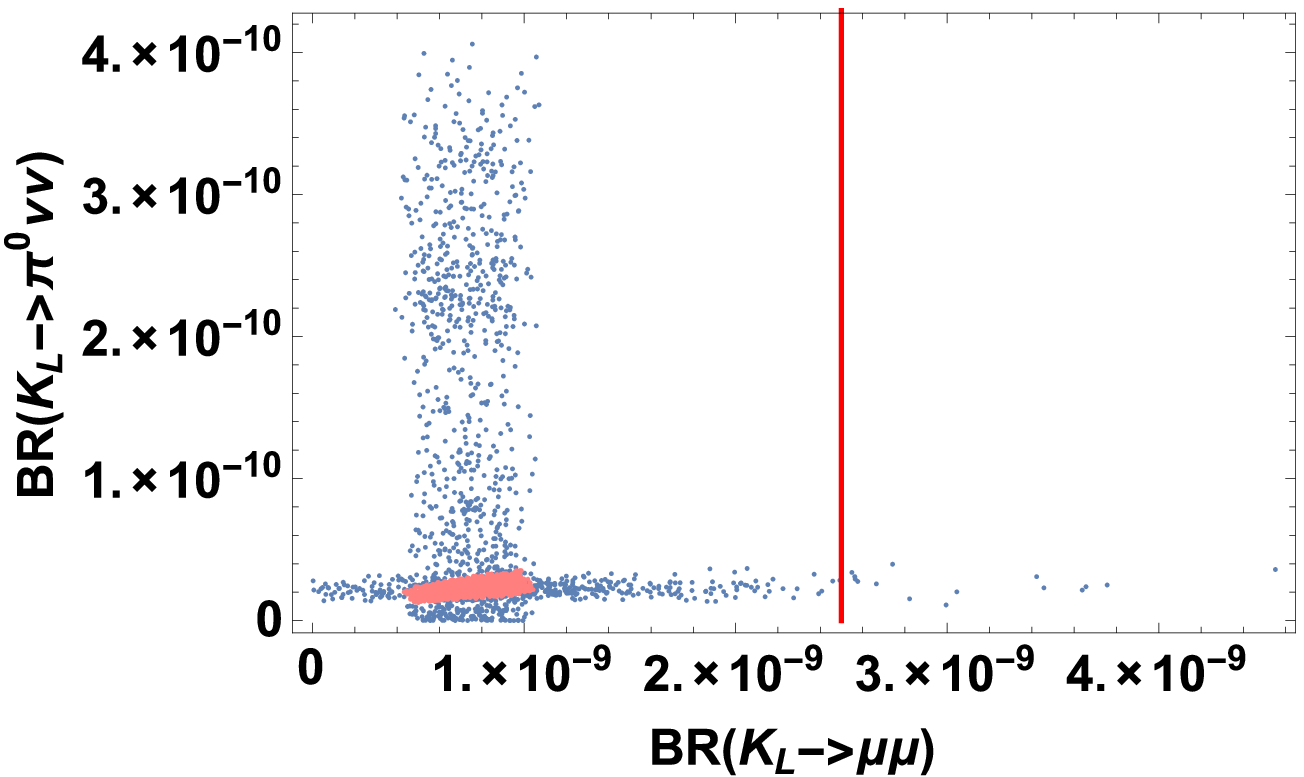}
\caption{The predicted ${\rm BR}(K_L \to \pi^0 \nu  {\bar \nu})$ versus
${\rm BR}(K_L\rightarrow  \mu^+\mu^-)$. 
The pink  denotes the SM  with $3\sigma$.
The  solid red line denotes the bound for the short-distance contribution.}
\end{minipage}
\hspace{5mm}
\begin{minipage}[]{0.48\linewidth}
\includegraphics[width=8cm]{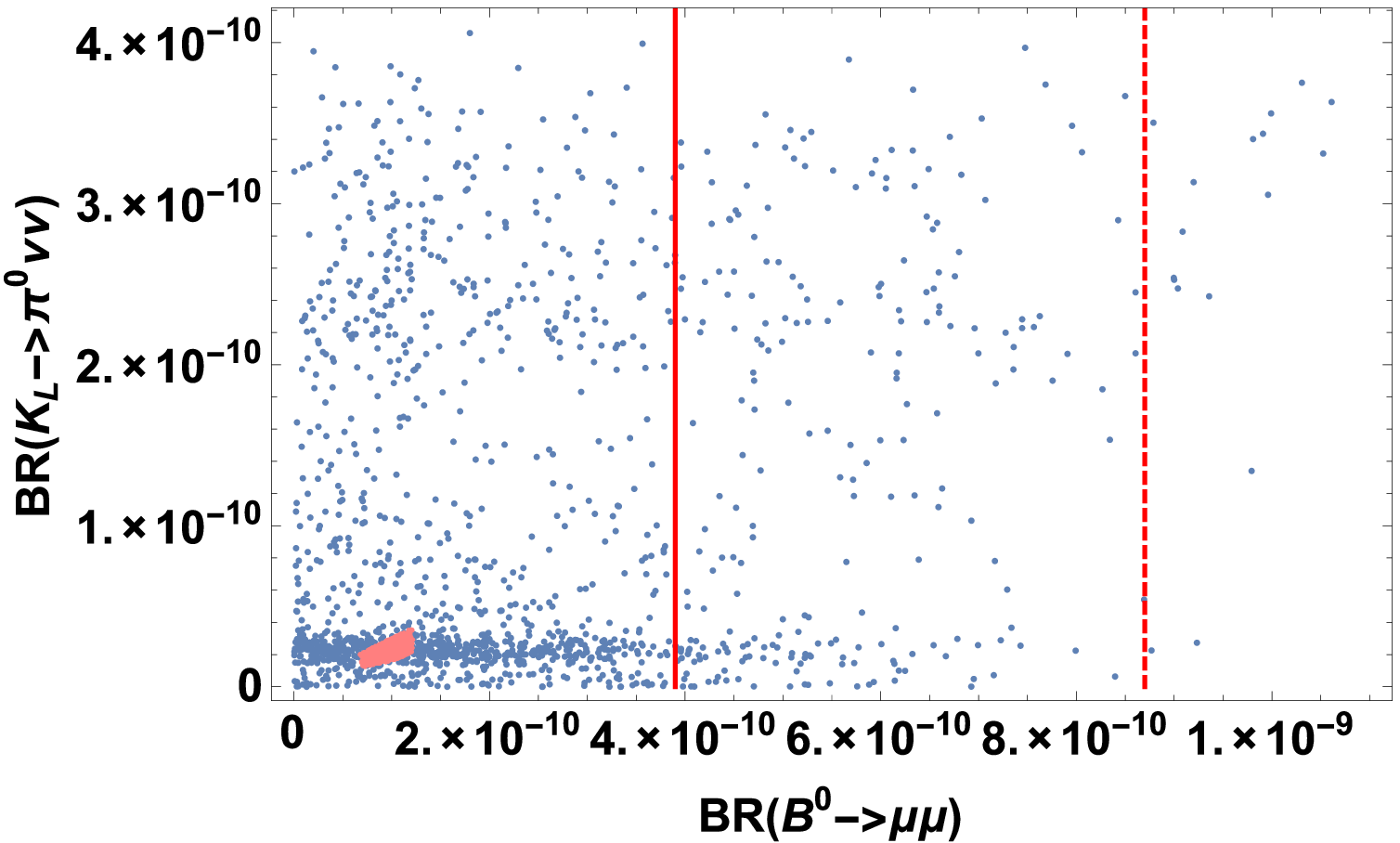}
\caption{The predicted ${\rm BR}(K_L \to \pi^0 \nu  {\bar \nu})$ versus
${\rm BR}(B^0\rightarrow \mu^+\mu^-)$.
The  solid red line denotes the central value of the  experimental data, and the  dashed one denotes the  experimental upper bound with $3\sigma$.
The pink  denotes the SM  with $3\sigma$. }
\end{minipage}
\end{figure}


\subsection{Results in the split-family  with $10$ TeV stop and sbottom}

  Let us discuss the case of the  the split-family of SUSY with $10$ TeV stop and sbottom,  
  where 1st and 2nd family squark masses  are around 2 TeV.
 The constraint of  $\epsilon_K$ is seriously tight  for 
the CP violating phases associated with the squark mixing  in the split-family SUSY model.
Moreover, the $|\Delta F|=2$ processes receive too large  contributions from the 
 the first and second squarks because they are relatively light, at ${\mathcal O}(1)$ TeV.
Actually,   $\Delta M_{K}$,  $\Delta M_{B^0}$
and $\Delta M_{B_s}$ are predicted as 
\begin{equation}
 \frac{\Delta M_K}{\Delta M_K(SM)}\simeq  400\ , \qquad
 \frac{\Delta M_{B^0}}{\Delta M_{B^0}(SM)}\simeq  50\ ,
\qquad \frac{\Delta M_{B_s}}{\Delta M_{B_s}(SM)}\simeq  3\ .
\end{equation}

In addition, the large left-right mixing generates large contributions to the $b \to s \gamma$ decay,
therefore, the left-right mixing angle  is severely constrained by experimental data
of $b \to s \gamma$.  
Therefore, it is impossible  to realize the  enhancement of
  ${\rm BR}(K_L \to \pi^0 \nu  {\bar \nu})$ in the split-family model
satisfying   constraints  of $|\Delta F|=1,2$  transitions in the kaon and the B meson systems.

\subsection{EDMs of neutron and   mercury}

Finally, we add a comment on
the electric dipole moments (EDMs) of the neutron and  the mercury (Hg), $d_n$ and $d_{Hg}$, which  arise through the chromo-EDM of the quarks, $d_q^C$
 due to  the gluino-squark mixing \cite{Pospelov:2000bw}-\cite{Fuyuto:2013gla}.
 If  both left-handed and right-handed mixing angles are taken to be large such as 
 $s_{13}^{dL}=s_{13}^{dR}\simeq 0.3$ or $s_{23}^{dL}=s_{23}^{dR}\simeq 0.3$ with the large left-right mixing, 
   $d_n$ and $d_{Hg}$ are predicted to be one and  two orders larger than the experimental upper bound, respectively \cite{PDG},
$|d_n| < 0.29 \times 10^{-25}{\rm e \cdot cm}$ and 
$|d_{Hg}| < 3.1 \times 10^{-29}{\rm e \cdot cm}$.

However, 
there  still  remains the freedom of phase parameters. 
For example, by tuning $\phi^{dL}_{i3}$ and $\phi^{dR}_{i3}$ $(i=1,2)$ under the constraint from $\epsilon_K$,
 we can suppress the  EDMs enough.
This tuning do not spoil our numerical results above.

\section{Summary and discussions}
\label{sec:summary}

In order to probe the SUSY    at  the $10 \ {\rm TeV}$ scale, we have studied
  the processes of $K_L \to \pi^0 \nu{\bar \nu}$ and $K^+ \to \pi^+ \nu{\bar \nu}$ 
combined with   the CP violating parameters $\epsilon_K$ and $\epsilon_K^\prime/\epsilon_K$.
The Z-penguin mediated by the chargino loop cannot enhance  $K_L \to \pi^0 \nu{\bar \nu}$ and
 $K^+ \to \pi^+ \nu{\bar \nu}$ 
 because the left-right mixing of the stop is constrained by the  $125$ GeV Higgs mass. 
On the other hand, the Z-penguin mediated by the gluino loop can enhance 
the  branching ratios of both $K_L \to \pi^0 \nu  {\bar \nu}$ and
 $K^+ \to \pi^+ \nu  {\bar \nu}$, 
 where the former increases more than $1.0 \times 10^{-10}$, 
 much larger than the SM prediction even if  the constraint of $\epsilon_K$ is imposed.
 Thus, the  $K_L \to \pi^0 \nu  {\bar \nu}$ and $K^+ \to \pi^+ \nu  {\bar \nu}$ decays
provide us very important information to probe the SUSY.
 
 It is remarkable that  the Z-penguin mediated by the gluino loop can
 enhance  simultaneously  $\epsilon^\prime_K/\epsilon_K$ and the branching ratio for
$K_L \to \pi^0 \nu  {\bar \nu}$.  While the estimated $\epsilon^\prime_K/\epsilon_K$ fits the observed value,
 the branching ratio  of $K_L \to \pi^0 \nu  {\bar \nu}$ increses  up to  $1.0 \times 10^{-10}$.
 
  We have also studied the decay rates of 
$K_L\rightarrow  \mu^+\mu^-$, $B^0\rightarrow \mu^+\mu^-$ and  $B_s\rightarrow \mu^+\mu^-$,
 which correlate  with the $K_L \to \pi^0 \nu  {\bar \nu}$ decay through the Z-penguin.
Especially, it is important to examine the $B^0\rightarrow \mu^+\mu^-$ decay carefully
since we can expect the enough  sensitivity of  the SUSY in this decay mode  at LHCb .


 We have also discussed them in the split-family model of SUSY, where 
   the third family of squarks/sleptons is heavy,  ${\cal O}(10)$ TeV,
while the first and second  ones of squarks/sleptons and the gauginos 
have relatively low masses of ${\cal O}(1)$ TeV.
 The constraint of  $\epsilon_K$ is much seriously tight  for 
the CP violating phases associated with the squark mixing  in the split-family SUSY model.
Moreover, the $|\Delta F|=2$ processes receive too large  contributions from
 the first and second family squarks because they are relatively light, at ${\mathcal O}(1)$ TeV.
Therefore, it is impossible  to realize the  enhancement of
  ${\rm BR}(K_L \to \pi^0 \nu  {\bar \nu})$ in the split-family model.

\vspace{0.2 cm}
\noindent
{\bf Acknowledgment}

We would like to thank 
Motoi Endo, Toru Goto and Satoshi Mishima
for useful discussions and important comments.
This work is supported by JSPS Grants-in-Aid for Scientific Research, No.25-5222.

\newpage
\appendix{}
\section*{Appendix A : Squark flavor mixing matrix}

The flavor mixing and CP violation  are induced through the quark-squark-gaugino and 
the lepton-slepton-gaugino couplings.
The Lagrangian of the gaugino-quark-squark interaction is written as 
\begin{equation}
\mathcal{L}_\text{int}(\tilde G q\tilde q)=-i\sqrt{2}g_{1,2,3}\sum _{\{ q\} }\widetilde q_i^*(T^a)
\overline{\widetilde{G}^a}\left [(\Gamma_L^{(q)})_{ij}{\bm L}
+(\Gamma_R^{(q)})_{ij}{\bm R}\right ]q_j+\text{H.c.}~,
\end{equation}
where $\widetilde{G}^a$ is the gaugino field,  $T^a$ is the generator of the gauge group,
 and ${\bm L}$, ${\bm R}$ are projection operators. 
The left-handed and right-handed mixing matrixes 
$\Gamma _{L}^{(q)}$ and $\Gamma _{R}^{(q)}$ diagonalizes 
the $6\times 6$ squark mass matrix  $M_{\tilde q}^2$ in the super-CKM basis to  
the mass eigenstate basis as follows:  
\begin{align}
M_{\tilde q}^2
&=
\Gamma^{(q)\dagger}\,\mathrm{diag}(m_{\tilde q}^2)\,\Gamma^{(q)}
=
\left(
\begin{array}{c c}
M_{LL}^2 & M_{LR}^2 \\[1.5mm]
M_{RL}^2 & M_{RR}^2
\end{array}
\right),
\end{align}
where $\Gamma^{(q)}$ is the $6\times 6$ unitary matrix, and  it is decomposed
into the $3\times 6$ matrices as $\Gamma ^{(q)}=(\Gamma _{L}^{(q)}, \ \Gamma _{R}^{(q))})$.
The squark mass matrix $M_{\tilde q}^2$ in the super-CKM basis is the same as that in the SLHA notation \cite{Allanach:2008qqSkands:2003cj}.
We write 
$\Gamma _{L, R}^{(q)}$  as follows:
\begin{align}
	\Gamma _{L}^{(q)}&=
	\begin{pmatrix}
	c_{13}^{qL} & 0 & s_{13}^{qL} e^{-i\phi_{13}^{qL}} c_{\theta^q_{LR}} & 0 & 0 & -s_{13}^{qL} e^{-i\phi_{13}^{qL}} s_{\theta^q_{LR}} e^{i \phi^q_{LR}} \\
	-s_{23}^{qL} s_{13}^{qL} e^{i(\phi_{13}^{qL}-\phi_{23}^{qL})} & c_{23}^{qL} & s_{23}^{qL}c_{13}^{qL}e^{-i\phi_{23}^{qL}}	c_{\theta^q_{LR}} & 0 & 0 & -s_{23}^{qL}c_{13}^{qL}e^{-i\phi_{23}^{qL}}s_{\theta^q_{LR}}e^{i\phi^q_{LR}} \\
	-s_{13}^{qL}c_{23}^{qL}e^{i\phi_{13}^{qL}} & -s_{23}^{qL}e^{i\phi_{23}^{qL}} &c_{13}^{qL}c_{23}^{qL}c_{\theta^q_{LR}} & 0 & 0 &  -c_{13}^{qL}c_{23}^{qL}s_{\theta^q_{LR}} e^{i\phi^q_{LR}}
\end{pmatrix}^T, \nonumber \\
\nonumber\\
\Gamma _{R}^{(q)}&=
\begin{pmatrix}
	0 & 0 & s_{13}^{qR} s_{\theta^q_{LR}} e^{-i \phi_{13}^{qR}} e^{-i\phi^q_{LR} } & c_{13}^{qR} & 0 & s_{13}^{qR} e^{-i \phi_{13}^{qR}} c_{\theta^q_{LR}} \\
	0 & 0 & s_{23}^{qR} c_{13}^{qR} s_{\theta^q_{LR}} e^{-i \phi_{23}^{qR}} e^{-i\phi^q_{LR} } & -s_{13}^{qR} s_{23}^{qR} e^{i(\phi_{13}^{qR}-\phi_{23}^{qR})} & c_{23}^{qR} & s_{23}^{qR} c_{13}^{qR} e^{-i \phi_{23}^{qR}} c_{\theta^q_{LR}} \\
0 & 0 & c_{13}^{qR} c_{23}^{qR} s_{\theta^q_{LR}} e^{-i\phi^q_{LR}} & -s_{13}^{qR} c_{23}^{qR} e^{i\phi_{13}^{qR}} & -s_{23}^{qR} e^{i \phi_{23}^{qR}} & 
c_{13}^{qR} c_{23}^{qR} c_{\theta^q_{LR}} \\
\end{pmatrix}^T,
\label{mixing}
\end{align} 
where we use abbreviations $c_{ij}^{qL,qR}=\cos\theta_{ij}^{qL,qR}$, $s_{ij}^{qL,qR}=\sin\theta_{ij}^{qL,qR}$,  $c_{\theta^q} =\cos \theta^q$ and $s_{\theta^q} =\sin \theta^q$ with 
$\theta_{ij}^{qL,qR}$ being the mixing angles between {\it i} th and {\it j} th familes of squarks.
In these mixing matrices, we take $s_{12}^{qL,qR}=0$. 

The 3 $\times$ 3 submatrix $M_{LR}^2$ is given as follows:
\begin{align}
M_{LR}^2 &=
(m_{\tilde{q}_{3, 1}}^2-m_{\tilde{q}_{3, 2}}^2)\cos\theta^{q_3}_{LR}\sin\theta^{q_3}_{LR}
e^{i\phi_{LR}^q}
\nonumber\\
&\hspace{10mm}\times
\left(
\begin{array}{ccc}
s_{13}^{qL} s_{13}^{qR} e^{i(\phi_{13}^{qL}-\phi_{13}^{qR})}
& c_{13}^{qR} s_{13}^{qL} s_{13}^{qR} e^{i(\phi_{13}^{qL}-\phi_{13}^{qR})}
& c_{13}^{qR} c_{23}^{qR} s_{13}^{qL} e^{i\phi_{13}^{qL}}
\\
c_{13}^{qL} s_{13}^{qR} s_{23}^{qL} e^{i(\phi_{23}^{qL}-\phi_{13}^{qR})}
& c_{13}^{qL} c_{13}^{qR} s_{23}^{qL} s_{23}^{qR} e^{i(\phi_{23}^{qL}-\phi_{23}^{qR})}
& c_{13}^{qL} c_{13}^{qR} s_{23}^{qL} c_{23}^{qR} e^{i\phi_{23}^{qL}}
\\
c_{13}^{qL} s_{13}^{qR} c_{23}^{qL} e^{-i\phi_{13}^{qR}}
& c_{13}^{qL} c_{13}^{qR} c_{23}^{qL} s_{23}^{qR} e^{-i\phi_{23}^{qR}}
& c_{13}^{qL} c_{13}^{qR} c_{23}^{qL} c_{23}^{qR}
\end{array}
\right).
\end{align}
The left-right mixing angles $\theta^q_{LR}$ are given approximately as 
\begin{equation}
	\theta^{b}_{LR}
	\simeq \frac{m_b(A_{33}^{d, *} - \mu \tan \beta)}{m_{\tilde b_L}^2-m_{\tilde b_R}^2} ,
	\qquad\qquad
	\theta^{t}_{LR}
	\simeq \frac{m_t(A_{33}^{u, *} - \mu \cot \beta)}{m_{\tilde t_L}^2-m_{\tilde t_R}^2} .
	\label{LRmixing}
\end{equation}

\section*{Appendix B :  Chargino and gluino interactions induced Z-penguin}

The Z -penguin amplitude mediated  by the chargino, $P_{\rm ZL}^{sd}(\chi^{\pm})$  in our basis \cite{GotoNote} 
is given as follows:
\begin{align}
P_{\rm ZL}^{sd}(\chi^{\pm})
&=\frac{g_2^2}{4m_W^2}
\sum_{\alpha,\beta.I,J}
(\Gamma _{CL}^{(d)\dagger})^I_{\alpha d}
(\Gamma _{CL}^{(d)})_J^{\beta s}
\Big\{\delta^J_I (U_+^\dagger)_\beta^1 (U_+)^\alpha_1  
\ [\log x^{\mu_0}_I + f_2(x^I_\alpha, x^I_\beta) ]   \nonumber \\
&-2 \delta^J_I (U_-^\dagger)_\beta^1 (U_-)^\alpha_1 \sqrt{x^I_\alpha x^I_\beta}f_1(x^I_\alpha, x^I_\beta)
-\delta^{\alpha}_{\beta}\left( \tilde{\Gamma}_L^{(u)}\right)^J_I f_2(x_I^{\alpha},x_J^{\alpha}) \Big\} \ ,
\label{PLZ}
\end{align}  
where
\begin{equation}
(\Gamma _{CL}^{(d)})_I^{\alpha q}\equiv 
( \Gamma _{L}^{(u)}  V_{\rm CKM})_I^q (U_+)^\alpha_1
+\frac{1}{g_2}( \Gamma _{R}^{(u)} \hat f_U V_{\rm CKM})_I^q (U_+)^\alpha_2 \ ,
\label{intCKM}
\end{equation}  
and
\begin{equation}
\left(\tilde{\Gamma}_L^{(u)} \right)_I^{ \ J}
\equiv
\left( \Gamma _{L}^{(u)} \Gamma _{L}^{(u)\dagger}
\right)_I^{ \ J} \ , 
\end{equation}  
with
 $q=s, d$,  $I=1-6$ for up-squarks,  and $\alpha=1,2$  for charginos.
 Here, $(U_\pm)^\alpha_i$ denote the mixing parameters between the wino and the higgsino.
 
 The right-handed Z  penguin one, $P_{\rm ZR}^{sd}(\chi^{\pm})$  
is also given simply  by replacements between $L$ and $R$, etc. \cite{GotoNote}.

The Z -penguin amplitude mediated the gluino, $P_{\rm ZL}^{sd}(\tilde g)$ \cite{GotoNote} 
is written as follows:
\begin{align}
P_{\rm ZL}^{sd}(\tilde g)
=-\frac{2}{3}\frac{g_3^2}{m_W^2}
\sum_{I,J}
(\Gamma _{GL}^{(d)\dagger})^I_d (\tilde\Gamma^{(d)})_I^{\ J}
(\Gamma _{GL}^{(d)})_{\ J}^s
f_2(x_I^{\tilde g},x_J^{\tilde g}) \ ,
\label{PLG}
\end{align} 
where 
\begin{equation}
\left(\tilde{\Gamma}_R^{(d)} \right)_I^{ \ J}
\equiv
\left( \Gamma _{R}^{(d)} \Gamma _{R}^{(d)\dagger}
\right)_I^{ \ J} \ . 
\end{equation} 
The right-handed Z  penguin $P_{\rm ZR}^{sd}(\tilde g)$  
is also given simply  by replacements between $L$ and $R$. 

\section*{Appendix C : Basic formulae}
\label{sec:}

\subsubsection*{C1 : $K^+ \to \pi^+ \nu {\bar \nu}$ and $K_L \to \pi^0 \nu  {\bar \nu}$}

The effective Hamiltonian for $K \to \pi \nu {\bar \nu}$ in the SM is given as \cite{Buras:1998raa}: 
\begin{align}
	{\mathcal H_{\rm eff}^{\rm SM}}
	=\frac{G_F}{\sqrt 2} \frac{2 \alpha}{\pi {\rm sin}^2\theta_W}
	\sum_{i=e, \mu, \tau} \left[ V_{cs}^*V_{cd} X_c + V_{ts}^*V_{td} X_t\right]
	\left( {\bar s_L}\gamma^{\mu} d_L \right)\left ({\bar \nu_L^i}\gamma_{\mu} \nu_L^i \right)+{\rm H.c.},
	\label{effH}
\end{align}
which is  induced by the box  and the Z-penguin mediated the W boson. 
The loop function $X_c$ denotes the charm-quark contribution of the  Z-penguin, and $X_t$ is the sum of the top-quark exchanges of the  box diagram and the  Z-penguin in Eq.(\ref{effH}).

Let us define the function $F$ as follows: 
\begin{align}
	F=V_{cs}^*V_{cd} X_c+V_{ts}^*V_{td} X_t \ .
\end{align}
The branching ratio of $K^+ \to \pi^+ \nu {\bar \nu}$ is given in terms of $F$.
Taking 
the ratio of it to the branching ratio of  $K^+ \to \pi^0 e^+ \nu$, which is the tree level transition, we obtain a simple form:
\begin{align}
	\frac{{\rm BR}(K^+ \to \pi^+ \nu {\bar \nu})}{{\rm BR}(K^+ \to \pi^0 e^+ \nu)}
	=\frac{2}{|V_{us}|^2} 
	\left(\frac{\alpha}{2\pi {\rm sin}^2\theta_W}\right)^2
	\sum_{i=e, \mu, \tau} |F|^2 .
	\label{BrKp}
\end{align}
Here the hadronic matrix element has been removed by using 
the fact that
the hadronic matrix element of $K^+ \to \pi^0 e^+ \nu$, 
which is well measured as ${\rm BR}(K^+ \to \pi^0 e^+ \nu)_{\rm exp} = (5.07 \pm 0.04)\times10^{-2}$ \cite{PDG},
 is related to the one of $K^+ \to \pi^+ \nu {\bar \nu}$ with the isospin symmetry:
\begin{align}
	\langle \pi^0 | \left( {\bar d_L}\gamma^{\mu} s_L \right) | {\bar K^0}\rangle
	&=\langle \pi^0 | \left( {\bar s_L}\gamma^{\mu} u_L \right) | K^+ \rangle , \\
	\langle \pi^+ | \left( {\bar s_L}\gamma^{\mu} d_L \right) | K^+\rangle
	&={\sqrt 2}\langle \pi^0 | \left( {\bar s_L}\gamma^{\mu} u_L \right) | K^+ \rangle .
\end{align}
Finally, the branching ratio for $K^+ \to \pi^+ \nu {\bar \nu}$ is expressed 
as follows:
\begin{align}
	{\rm BR}(K^+ \to \pi^+ \nu {\bar \nu})
	=3 \kappa |F|^2,  
	\qquad
	\kappa
	=\frac{2}{|V_{us}|^2} r_{K^+} 
	\left(\frac{\alpha}{2\pi {\rm sin}^2\theta_W}\right)^2 {\rm BR}(K^+ \to \pi^0 e^+ \nu),
	\label{BRKpPipnunu}
\end{align}
where $r_{K^+}$ is the isospin breaking correction
 between $K^+ \to \pi^+ \nu {\bar \nu}$ and $K^+ \to \pi^0 e^+ \nu$ 
\cite{Marciano:1996wy,Mescia:2007kn}, and 
the factor 3 comes from the sum of three neutrino flavors.
It is noticed that the branching ratio for $K^+ \to \pi^+ \nu {\bar \nu}$ depends on both the real and imaginary parts of  $F$.   

For the $K_L \to \pi^0 \nu  {\bar \nu}$ decay, 
the $K^0-{\bar K^0}$ mixing should be taken account, and one obtains
\begin{align}
	A(&K_L \to \pi^0 \nu  {\bar \nu})
	=\frac{G_F}{\sqrt 2} \frac{2 \alpha}{\pi {\rm sin}^2\theta_W}
	\left ({\bar \nu_L^i}\gamma_{\mu} \nu_L^i \right)
	\langle \pi^0| 
	\left[ F ({\bar s_L}\gamma_{\mu} d_L)+F^* ({\bar d_L}\gamma_{\mu} s_L) \right]
	 | K_L\rangle \nonumber \\ 
	&=\frac{G_F}{\sqrt 2} \frac{2 \alpha}{\pi {\rm sin}^2\theta_W}
	\left ({\bar \nu_L^i}\gamma_{\mu} \nu_L^i \right)
	\frac{1}{\sqrt 2}
	\left[ F(1+\bar{\epsilon})\langle \pi^0| ({\bar s_L}\gamma_{\mu} d_L) | K^0\rangle
	+F^*(1-\bar{\epsilon})\langle \pi^0| ({\bar d_L}\gamma_{\mu} s_L) | \bar K^0\rangle
	\right] \nonumber \\
	&\simeq \frac{G_F}{\sqrt 2} \frac{2 \alpha}{\pi {\rm sin}^2\theta_W}
	\left ({\bar \nu_L^i}\gamma_{\mu} \nu_L^i \right)
	\frac{1}{\sqrt 2} \
	2 \ i \ {\rm Im}F \
	\langle \pi^0| ({\bar s_L}\gamma_{\mu} d_L) | K^0\rangle ,
\label{A(KLpi0nun)}
\end{align}
where  we use
\begin{align}
	 | K_L\rangle = \frac{1}{\sqrt 2}
	\left[ (1+\bar{\epsilon}) | K^0\rangle
	+(1-\bar{\epsilon}) | K^0 \rangle \right] ,
\end{align}
with  
\begin{align}
	{\rm CP} | K^0\rangle
	=-| \bar K^0\rangle, \quad
	\langle \pi^0| ({\bar d_L}\gamma_{\mu} s_L) | \bar K^0\rangle
	=-\langle \pi^0| ({\bar s_L}\gamma_{\mu} d_L) | K^0\rangle .
\end{align}
We neglect the CP violation in $K^0-{\bar K^0}$ mixing, ${\bar \epsilon}$, due to its smallness $|{\bar \epsilon}| \sim 10^{-3}$.
Taking the ratio between the branching ratios of  $K^+ \to \pi^0 e^+ \nu$
and  $K_L \to \pi^0 \nu  {\bar \nu}$, we have the simple form:
\begin{align}
	\frac{{\rm BR}(K_L \to \pi^0 \nu  {\bar \nu})}{{\rm BR}(K^+ \to \pi^0 e^+ \nu)}
	=\frac{2}{|V_{us}|^2} 
	\left(\frac{\alpha}{2\pi {\rm sin}^2\theta_W}\right)^2 \frac{\tau(K_L)}{\tau(K^+)} 
	\sum_{i=e, \mu, \tau} ({\rm Im} F)^2 .
\end{align}
Therefore,  the branching ratio of  $K_L \to \pi^0 \nu  {\bar \nu}$ is given as follows:
\begin{align}
	{\rm BR}(K_L \to \pi^0 \nu  {\bar \nu})
	=3 \kappa \cdot \frac{r_{K_L}}{r_{K^+}}\frac{\tau (K_L)}{\tau (K^+)} ({\rm Im} F)^2 , 
\label{BRKLPi0nunu}
\end{align}
where $r_{K_L}$ denotes the isospin breaking effect 
\cite{Marciano:1996wy,Mescia:2007kn}.
It is remarked that the branching ratio of $K_L \to \pi^0 \nu  {\bar \nu}$ depends on the imaginary part of $F$.


The effective Hamiltonian in Eq.(\ref{effH}) is modified due to   new box diagrams and penguin diagrams  induced by SUSY particles. Then, the effective Lagrangian  is given as 
\begin{align}
	{\mathcal L_{\rm eff}}
	=\sum_{i, j=e, \mu, \tau} \left[ C_{\rm VLL}^{ij} \left( {\bar s_L}\gamma^{\mu} d_L \right)
	+C_{\rm VRL}^{ij} \left({\bar s_R}\gamma^{\mu} d_R \right)\right]\left ({\bar \nu_L^i}\gamma_{\mu} \nu_L^j \right)+{\rm H.c.} \ ,
\end{align}
where $i$ and $j$ are the indices of the flavor of the neutrino final state.
Here, $C_{\rm VLL, VRL}^{ij}$ is the sum of the box contribution and the Z-penguin  one:
\begin{align}
	C_{\rm VLL}^{ij}
	=-B_{\rm VLL}^{sdij}-\frac{\alpha_2}{4 \pi}Q_{ZL}^{(\nu)}P_{\rm ZL}^{sd} \delta^{ij} \ , \ \
	C_{\rm VRL}^{ij}
	=-B_{\rm VRL}^{sdij}-\frac{\alpha_2}{4 \pi}Q_{ZL}^{(\nu)}P_{\rm ZR}^{sd} \delta^{ij} \ ,
\end{align}
where the weak neutral-current coupling $Q_{ZL}^{(\nu)}=1/2$,
and $B_{\rm VL(R)L}^{sdij}$ and  $P_{\rm ZL(R)}^{sd}$ denote
 the box  contribution  and the Z-penguin contribution, respectively. 
 The $V$, $L$ and  $R$ denote the vector coupling, the left-handed one and the right-handed one,
 respectively.
In  addition to the W boson contribution,  there are the gluino $\tilde{g}$, 
the chargino $\chi^{\pm}$ and the neutralino $\chi^{0}$ mediated ones.

The branching ratios of $K^+ \to \pi^+ \nu {\bar \nu}$ and $K_L \to \pi^0 \nu  {\bar \nu}$  are obtained by replacing internal effect $F$ in Eqs. (\ref{BRKpPipnunu}) and 
(\ref{BRKLPi0nunu}) to $C_{\rm VLL}^{ij}+C_{\rm  VRL}^{ij}$  as follows:
\begin{align}
	{\rm BR}(K^+ \to \pi^+ \nu {\bar \nu})
	&=\kappa 
	\sum_{i=e, \mu, \tau} |C_{\rm VLL}^{ij}+C_{\rm VRL}^{ij}|^2 \ ,\\
	{\rm BR}(K_L \to \pi^0 \nu  {\bar \nu})
	&=\kappa \cdot \frac{r_{K_L}}{r_{K^+}}\frac{\tau (K_L)}{\tau (K^+)} 
	\sum_{i=e, \mu, \tau} |{\rm Im} (C_{\rm VLL}^{ij}+C_{\rm VRL}^{ij})|^2 \  .
\end{align}

\subsection*{C2 : $B_s\rightarrow \mu^+ \mu^-$, $B^0\rightarrow \mu^+ \mu^-$
and $K_L\rightarrow \mu^+ \mu^-$}
The Z-penguin process appears in  $B_s\rightarrow \mu^+ \mu^-$ and $B^0\rightarrow \mu^+ \mu^-$ 
decays. 
We show the branching ratio for  $B_s\rightarrow \mu^+ \mu^-$,
which includes  the Z-penguin amplitude \cite{GotoNote}:
\begin{align}
{\rm BR}(B_s\rightarrow \mu^+ \mu^-)
 &=\tau_{B_s}
 \frac{m_{B_s}^3 f_{B_s}^2}{16 \pi} \left( \frac{\alpha}{4 \pi} \right)^2
 \frac{m_{\mu}^2}{m_{B_s}^2}
 \sqrt{1-\frac{4m_\mu^2}{m_{B_s}^2}} \left |  C_{VRA}^{(\mu)}-C_{VLA}^{(\mu)}\right |^2 \ ,
\end{align} 
where
\begin{equation}
\frac{\alpha}{4\pi} C_{VLA}^{(\mu)}=-B_{VLL}^{(bs\mu\mu)}(SM)-\frac{\alpha_2}{4\pi} \frac{1}{4} P_{ZL}^{bs}\ ,
\quad \quad
\frac{\alpha}{4\pi} C_{VRA}^{(\mu)}=-\frac{\alpha_2}{4\pi} \frac{1}{4} P_{ZR}^{bs}\ .
\end{equation} 
We include the  box diagram only for the SM, which is 
\begin{equation}
B_{VLL}^{(bs\mu\mu)}(SM)=-\frac{\alpha_2}{4\pi} \frac{g_2^2}{2m_W^2}  V_{tb}V_{ts}^* B_0(x_t).
\end{equation} 
On  the other hand, the SM component of the Z-penguin amplitude is
\begin{equation}
P_{ZL}^{bs}=\frac{g_2^2}{2m_W^2}  V_{tb}V_{ts}^* \times 4C_0(x_t),
\end{equation} 
where $B_0(x_t)$ and $C_0(x_t)$  are well known loop-functions depending on $x_t=m_t^2/m_W^2$.
We have neglected other amplitudes such as the  Higgs mediated scalar amplitude
since we focus on  NP in the Z-penguin process.

The branching ratio of $B^0\rightarrow \mu^+ \mu^-$ 
is  given in the similar expression.
For  the $K_L \rightarrow \mu^+ \mu^-$  decay, 
its branching ratio is given as follows \cite{Gorbahn:2006bm} :
\begin{align}
{\rm BR} (K_L \to \mu^+ \mu^{-})_{\rm SD}
&=
\kappa_{\mu} \left[ 
\frac{{\rm Re} \lambda_t}{\lambda^5} Y(x_t)
+ \frac{{\rm Re} \lambda_c}{\lambda} P_c
\right]^2 , \\
\kappa_{\mu}
&= (2.009 \pm 0.017) \times 10^{-9}
\left( \frac{\lambda}{0.225}\right)^8 ,
\end{align}
where $\lambda$ is the Wolfenstein parameter, $\lambda_{i}= V_{is}^* V_{id}$
and the charm-quark contribution $P_c$ is calculated in NNLO ; $P_c = 0.115 \pm 0.018$, 
and $Y$ is the same as in eq.(7).
We use its SM value as $Y(x_t) = 0.950 \pm 0.049, \ (x_t \equiv m_t^2/M_W^2)$.

\newpage

\end{document}